%% file: DTGravity-v5.tex
\documentclass[12pt,letterpaper]{JHEP3}

\usepackage[latin1]{inputenc} 
\usepackage{bm}
\usepackage{amsmath,amssymb,graphicx,mathrsfs,epsfig}
\usepackage{caption}
\usepackage{cite}
\usepackage{ulem}
\newcommand{\eqn}[1]{Eq.~(\ref{#1})}
\newcommand{\eqns}[2]{Eqs.~(\ref{#1}), (\ref{#2})}
\newcommand{\reference}[1]{Ref.~\cite{#1}}

\def\bcal{{\cal B}}
\def\dcal{{\cal D}}
\def\ecal{{\cal E}}
\def\gcal{{\cal G}}
\def\lcal{{\cal L}}
\def\ab{{\alpha\beta}}
\def\kl{{\kappa\lambda}}
\def\mn{{\mu\nu}}
\def\viz{viz.}
\def\ie{{i.\,e.}}
\def\eg{{e.\,g.}}
\def\half{\frac12}
\def\pa{{\partial}}
\def\vevof#1{\left\langle #1 \right\rangle}
\def\vev{vacuum expectation value}

\input{mylatex}

\title{Naturalness and Dimensional Transmutation in Classically Scale-Invariant Gravity\vskip-5mm}

\author{Martin B Einhorn$^{1\thanks{Current address.}\,\,,\,2}$, 
D R Timothy Jones$^{1,\, 3\thanks{Current address.}}$\\
$^{1}$Kavli Institute for Theoretical Physics,
University of California, Santa Barbara
CA 93106-4030, USA\\
$^{2}$Michigan Center for Theoretical Physics, 
University of Michigan, Ann Arbor, MI 48109-1040, USA\\
$^{3}$Dept. of Mathematical Sciences,
University of Liverpool, Liverpool L69 3BX, UK\\
E-mail:  \email{meinhorn@umich.edu, drtj@liv.ac.uk}}

\abstract{
We discuss the nature of quantum field theories involving gravity that
are classically scale-invariant. We show that gravitational radiative
corrections are  crucial in the determination of  the nature of the
vacuum state in such theories, which are renormalisable, technically
natural, and can be asymptotically free in all dimensionless couplings. 
In the pure gravity case, we discuss the role of the Gauss-Bonnet
term, and we find that Dimensional Transmutation (DT) \`a la
Coleman-Weinberg leads to extrema of the effective action corresponding
to nonzero values of the  curvature, but such that these extrema are  
local maxima. In even the simplest extension of the  theory to include 
scalar fields, we show that the same phenomenon can lead to extrema
that are local minima of the effective action, with both non-zero   
curvature  and non-zero scalar vacuum expectation values, leading to
spontaneous generation of the Planck mass. Although we find an
asymptotically free (AF) fixed point exists, 
unfortunately, no running of the couplings connect the
region of DT to the basin of attraction of the AF fixed point.
We also find there remains a flat direction for one of 
the conformal modes. We suggest that in more realistic models 
AF and DT could be compatible, and that the same
scalar vacuum expectation values could be responsible both for DT and for
spontaneous breaking of a Grand Unified gauge group.}


\begin{document}
\pagebreak
\section{Introduction}\label{sec:intro}

\paragraph{ }
Classically scale-invariant and conformally invariant models have
attracted great interest in quantum field theory (QFT) for a very long
time in a variety of contexts, both phenomenological and theoretical. 
These symmetries are anomalous in QFT in four-dimensions except in rare
circumstances, such as N=4 supersymmetric Yang-Mills theory. This
symmetry breaking is inherent in the renormalization process, leading to
the concept of scale-dependent or running coupling constants. 

Why, then, should one be interested in classically scale-invariant
theories? One reason is that, in the search for the origin of masses,
dimensional transmutation (DT) is a mechanism that can ``explain" the
appearance of a mass scale from an otherwise massless theory and can
lead to definite relationships among masses that are not simply the
consequence of internal symmetries. A second motivation  is also that
classically scale-invariant models that include the metric tensor are 
renormalizable~\cite{Stelle:1976gc}, and their coupling constants are
asymptotically free (AF) or asymptotically finite,  at least for some
range of parameters\footnote{\reference{Avramidi:2000pia}  contains a
detailed review of higher-order gravity.  (This is essentially a
reproduction of the author's 1986 PhD thesis [hep-th/9510140].) 
\reference{Buchbinder:1992rb} provides a comprehensive overview by some
of the pioneers in the field.  Unfortunately, this contains numerous
typographical  errors in the equations associated with
$R^2$-gravity.}~\cite{Fradkin:1981hx,  Fradkin:1981iu, Avramidi:1985ki,
Avramidi:2000pia}.  In principle, such models may
provide an ultraviolet (UV) completion of Einstein gravity. They offer
the prospect of generating the Planck mass $M_P$ dynamically, with
consistent physics at energies above $M_P.$ Even if not the final word,
they may provide a perturbatively calculable framework within which some
of the puzzles associated with quantum gravity may be given definite
answers. Needless to say, such models may  also be very important for
understanding the very early universe,  especially if inflation is an
ingredient.

A third motivation is that such theories retain a legacy of their
classical scale invariance inasmuch as  their symmetry-breaking is
``soft"~\cite{Bardeen:1995kv, Foot:2007iy, Altmannshofer:2014vra},  \ie,
the masses do not suffer from naturalness issues associated with
power-law divergences~\cite{'tHooft:1979bh}. 

In the simplest case of massless, scalar $\lambda\phi^4$ theory in flat
spacetime, the running of the coupling $\lambda(\mu)$ insures that, as
$\mu\to0,$ in fact $\lambda\to0$, and the model approaches a free field
theory. It also suggests that as $\mu\to\infty,$ $\lambda(\mu)$ becomes
large, so that, above some sufficiently high energy $\Lambda$, the
theory becomes strongly coupled. This is frequently interpreted as a
sign that, at scales above $\Lambda$ (often associated with the term
``Landau pole"), the theory is not simply strongly coupled but
incomplete or inconsistent.  In other cases, such as Yang-Mills (Y-M)
theory or massless QCD, the theory becomes AF at high momentum scales
$\mu$, with gauge coupling $g(\mu)\to0$ as $\mu\to\infty,$ but strongly
coupled below some low-energy scale $\Lambda$. Although not yet
rigorously proven, it is firmly believed that the result is gluon
condensation or quark confinement, \ie, unlike perturbation theory in
which the quanta are massless, the true spectrum of the theory has
massive particles whose mass scale is determined by where the effective
interaction strength, characterized by $\alpha(\mu)=g^2(\mu)/4\pi$,
becomes sufficiently large (typically, $\alpha(\mu)\sim 1.$) We shall
refer to this generically as dimensional transmutation (DT) due to
strong interactions.

Finally, there is a third possibility, first discussed by Coleman and
Weinberg (CW)~\cite{Coleman:1973jx}, in which a classically scale-invariant
theory generates a mass scale $\Lambda$ at
which a specific relationship among multiple couplings obtains. In the
case of scalar electrodynamics, this occurs for
$\lambda(\Lambda)\sim\alpha(\Lambda)^2$, which can be at weak coupling
where perturbation theory may still be a good approximation. This has
been called ``dimensional transmutation," whereby a massless theory
with two or more couplings can be described in terms of a mass scale $\Lambda$
and a single coupling $\alpha(\mu)$ together with a relation that determines
the second coupling $\lambda(\Lambda)$ at the specific scale $\Lambda$.
 If necessary, to distinguish this case from the strong-coupling
mechanism characteristic of theories like QCD, we shall refer to
this as perturbative or weak-coupling DT. 

Among classically scale-invariant theories is higher-order gravity, 
often referred to as $R^2$-gravity, described by a ``higher-order"
action such as\footnote{It is convenient, although probably not
necessary, to work with the Euclidean form of the QFT, and we shall do
so throughout this paper. For Einstein gravity, the Euclidean Path
Integral is not well-understood\cite{Carlip:2001wq}, and some of the
same issues would apply to $R^2$ theories~\cite{Barth:1983hb}. If
the spacetime manifold has boundaries, one needs to supplement this action 
integral~\cite{Gibbons:1976ue, Barth:1984jb}, but these will not be relevant to 
our applications.}
\beq
\label{eq:lho1}
S_{ho}=\int d^4x \sqrt{g} \left[ \frac{1}{2\alpha}C_{\kl\mn}^2+ \frac{1}{3\beta}R^2+\frac{2}{\gamma}{R}_\mn^2 \right],
\eeq
where $C_{\kl\mn}$ is the Weyl tensor, $R_\mn$ is the Ricci tensor, and $R$ the Ricci scalar. 
These are the maximum number of scalars of dimension four that can be formed from the Riemann-curvature tensor 
$R_{\kl\mn}$. The three coupling constants $\alpha,\beta,\gamma$ are dimensionless. 
As mentioned earlier,
with a propagator behaving as $1/q^4$, this theory has been shown to be
renormalizable and asymptotically free, with or without the addition of a
linear term in $R$, a cosmological constant, or, with some weak limitations,
with the inclusion of matter.

The low-energy behaviour of this theory is not well understood. If a
linear (Einstein-Hilbert) term $M^2R$ is explicitly added, the perturbative
spectrum in flat background has a massive scalar, a massless graviton,
and a massive, spin-two ghost~\cite{Tomboulis:1996cy}. For this reason,
the theory is often thought to violate unitarity; it seems as if this
model is just a clever way of embedding a Pauli-Villars ghost in a
manner consistent with general covariance and achieving its
renormalizability in an unphysical way. This interpretation of the
classically scale invariant theory leaves room for doubt for
several reasons.  Perhaps it is
simply the way in which mass was introduced that is at fault. The
static potential associated with a $1/q^4$ propagator is proportional to
distance, $|\vec{x}|$, so, taken at face value, this would be a
confining theory! The same conclusion is also suggested by the running
of the gravitational couplings. The complement of their being
asymptotically free (AF) is that they grow as the renormalization scale $\mu$ 
decreases, so one would expect the theory to become strongly coupled 
at lower energy scales. It seems unlikely that the resulting theory at
large distances would look anything like general relativity, quite aside
from whether or not it satisfies unitarity. Finally, the spectrum in a flat 
space background may not be relevant to theories having a curved
spacetime background.  It is notoriously difficult to determine the 
candidate no-particle states (or vacua) when gravity is included.

Another possibility is that the theory 
(\!{\it without\/} an explicit Einstein-Hilbert term) undergoes DT of the
CW type discussed above, where gravitational couplings 
play the role of the electromagnetic coupling in scalar electrodynamics~\cite{Fradkin:1981iu}.  
If that occurs and the couplings are weak,
then it should be possible to infer the properties of the theory at that
scale and of the effective field theory below. We shall show that,
assuming maximally symmetric spacetime, DT can occur for weak couplings
in $R^2$-gravity (\eqn{eq:lho1}), but the extrema are not locally stable and 
cannot be assumed to be the true no-particle or vacuum state.

If this theory is to look at all like Einstein gravity at low energies, it seems to be
necessary to include matter.  In order for $R^2$-gravity 
to remain natural, the matter field action must not only
describe a renormalizable theory but also be classically scale invariant. 
This is automatically true for gauge bosons, but it is a strong constraint 
on scalars and fermions. Previous such attempts have been plagued 
by an effective action that contains an imaginary part, 
(reviewed, \eg, in~\reference{Odintsov:1989gz},)
suggesting that such models become unstable.  
We shall see that is not the case here.
   
The simplest form of matter would be to add a real
scalar field to \eqn{eq:lho1} in a classically scale-invariant manner
\beq\label{eq:jrealscalar} 
S_m=\int d^4x \sqrt{g}\left[ \half
(\nabla\phi)^2+\frac{\lambda}{4}\phi^4-\frac{\xi\phi^2}{2}R \right].
\eeq 
The non-minimal coupling $\xi$ is required for renormalizability. 
Perturbatively, of course, one might expect $\vevof{\phi}=0,$ as in the
purely scalar theory. However, if $R^2$-gravity plays the role of
electrodynamics in the Coleman-Weinberg model,  
DT may occur for some relation among the various couplings. If so, and
$\vevof{\phi}=v \neq 0$, then $8\pi \xi v^2\equiv M_P^2$ would correspond
to the Planck mass $M_P$ (assuming that $\xi>0$). Below this scale, the
theory would look very much like ordinary general relativity. We shall
show that DT can in fact occur and that the effective action in this model
does not have an imaginary part.  
 We find that this model has several fixed points, one of which
is indeed AF.  Unfortunately, even though the gravitational couplings $a, b$ are AF, 
the basin of attraction of this AF fixed point does not include 
the range of matter couplings at which these minima occur.
(One may hope that this disappointing result is model dependent and that
more realistic models including, \eg, non-Abelian gauge fields and fermions,
might not encounter such an obstruction).
Nevertheless, thinking of $S_{ho}+S_m$ 
as an effective field theory, these 
minima are candidates for stable vacua at the Planck
mass scale and below.  Whether they are also 
unitary theories has not been determined, 
although we shall discuss the issue further.

The outline of this paper is as follows: In the next section, we discuss
the theory defined by the action \eqn{eq:lho1} and its 
renormalization in terms of essentially two coupling constants. 
In connection with this 
we explain the role of the Gauss-Bonnet term, and remark on 
the relationship of its renormalization and a possible $a$-theorem. Then, in
Section~\ref{sec:natural}, we describe the nature of scale symmetry
breaking in QFT and its implications for naturalness. In
Section~\ref{sec:dtgravity}, we review the effective action in
$R^2$-gravity, including the one-loop beta-functions for the couplings
and their asymptotic freedom (AF). We show that the beta-functions for
the couplings determine the form of the one-loop, $O(\hbar)$, correction
to the effective action and investigate the possibility of DT at its
extrema. We derive a (new) formula for the local curvature in order to
determine whether an extremum is a (local) maximum, minimum, or
saddle-point. An interesting aspect of this development is that, even
though the curvature is $O(\hbar^2),$ it is determined entirely by the
one-loop corrections. In Section~\ref{sec:realscalar}, we extend this
formalism to the model with a massless, real scalar field, showing how
DT may arise and discuss the low-energy effective field theory. Although
we use the Jordan frame for the most part, we also discuss this model
from the point of view of the Einstein frame. In Section~\ref{sec:morematter}, we
briefly discuss extending the model to include the Standard Model fields.
In Section~\ref{sec:constraints}, we discuss constraints on the  coupling
constants in order to expect theories of this sort to make sense both at
the highest possible scales as well as at and below the DT scale.
Finally, in Section~\ref{sec:conclusions}, we conclude with a
discuss of open questions and future applications. In five 
appendices, we review some topics that bear on our work in an effort
to make this paper more self-contained, \viz, the
Gauss-Bonnet relation, the background field method, 
the definition of scale invariance, constraints on the couplings required 
for stability, and the one-loop beta functions for models of this type. 

\section{The Action for Pure $R^2$ ``Gravity"}\label{sec:hogravity}

\paragraph{ }
Because we are interested in classically\footnote{By ``classical," we
simply mean the tree approximation in terms of renormalized couplings
and fields  associated with some conveniently chosen scale.} scale
invariant theories in four dimensions, the action for pure gravity will
contain the quadratic invariants given in \eqn{eq:lho1} However, this is
not the action that has been the  starting point for analyses of this
theory~\cite{Stelle:1976gc,Fradkin:1981hx,Fradkin:1981iu,Avramidi:1985ki}. 
This is because of the Gauss-Bonnet relation,
which in differential form may be expressed as 
\beq\label{eq:g-b}
 R^*R^*=C_{\kl\mn}^{\, 2}-2\widehat{R}_\mn^{\, 2}+\frac{1}{6}R^{\,2} \equiv G,
\eeq 
where $\widehat{R}_\mn\equiv R_\mn-g_\mn R/4$ is the traceless Ricci tensor.
The properties of the ``topological" term $ R^*R^*$ and its
relation in integral form to the Euler characteristic are summarized in
Appendix~\ref{sec:g-b}. For our purposes, it is sufficient to know
that it can be written as the divergence of a current,
$R^*R^*=\nabla_\mu B^\mu.$  
As a result, the variation of its contribution to the action vanishes identically 
\beq\label{eq:varindex} 
\frac{\delta }{\delta g_\mn}\int\!\! d^4x
\sqrt{g}\ G=0. 
\eeq 
This property is closely related to the validity
of the Bianchi identities. Although special to four dimensions, these
act like another symmetry that reduces the number of independent
couplings.  The action \eqn{eq:lho1} can be rewritten, for example, as 
\beq\label{eq:lho2} 
S_{ho}=\int d^4x \sqrt{g} \left[
\frac{1}{2a}C_{\kl\mn}^2+ \frac{1}{3b}R^2+\varepsilon G \right]. 
\eeq
According to \eqn{eq:varindex}, the last term contributes nothing to the
variation of the action, so one might think it could be discarded
altogether. When formulating the Feynman rules in four-dimensions, 
it is certainly irrelevant so that, in fact, this theory would appear to be
renormalizable in terms of two coupling constants only ($a,b$). 
However, the theory without the $\varepsilon G$ term is not 
multiplicatively renormalizable\footnote{This
has nothing to do with the regularization chosen.  It is
the fact that there are three independent scalar quadratic invariants,
 given in \eqn{eq:lho1}, and divergences occur proportional 
 to each of them. It 
would be equally true using a regularization scheme 
operating within four-dimensions. This is discussed further
below and in Appendix~\ref{sec:bfm}.}.
Because of its relative simplicity and manifest gauge invariance, 
the regularization scheme usually 
chosen is dimensional regularization (DREG).  This confuses the issue further because,
for dimension $n\ne4,$ the operator 
$\sqrt{g}\,G$ cannot be expressed as a 
total derivative, nor can any dimension-dependent linear combination  of
the three renormalized operators~\cite{Capper:1979pr}. 
One might be tempted to conclude that, like scale
invariance or chiral symmetry, the Gauss-Bonnet relation was
anomalous or at least inconsistent with DREG~\cite{Brunini:1993qz}.
Fortunately, it is enough to extend 
$\sqrt{g}(C_{\kl\mn}^2-2\widehat{R}_\mn^2+R^2/6)$ to $n$-dimensions,
which is possible.  Any definition for continuous $n$ that reduces 
to this linear combination as $n\to4$ should suffice.  This enables 
the definition of renormalized operators and couplings in
four dimensions, at which point, one may then rewrite 
$G=R^*R^*=\nabla_\mu B^\mu$ locally, using the special 
properties of the curvature tensor in four dimensions, 
such as the Bianchi identities.

Nevertheless, the extension of \eqn{eq:varindex} to $n$-dimensions will not be correct, so
one would think that one needs to include $G$ in constructing the Feynman rules in $n$-dimensions, 
adding further complications to renormalization of the theory.
In fact, this obstacle has been circumvented by previous 
authors~\cite{Stelle:1976gc, Fradkin:1981hx,Fradkin:1981iu,Avramidi:1985ki}.  
In practice, this has been accomplished as follows:
Ignoring $\varepsilon G$ when determining the Feynman rules,
one finds that the theory is not multiplicatively renormalizable unless
one includes counterterms for $\varepsilon G$ as well.  Since these divergences
(up to finite local counterterms) determine the beta-functions, this would imply
that $\beta_\varepsilon$ is a function only of the remaining couplings, $a,b$.
If so, then it must be the case that
\beq\label{eq:rgi}
\beta_{\varepsilon}=\frac{\pa\varepsilon}{\pa a}\beta_a(a,b)+
\frac{\pa\varepsilon}{\pa b}\beta_b(a,b),
\eeq
to all orders in perturbation theory.
This is a nontrivial statement about the renormalized couplings
in four-dimensions.  Among other things, it implies that 
there must then be a tree-level contribution to 
$\varepsilon(a,b)$ as well.
In this theory, because \eqn{eq:varindex}
is correct in four dimensions, it is possible to solve \eqn{eq:rgi}
order-by-order in perturbation theory. 
In a separate publication~\cite{Einhorn:2014bka}, we
prove this is possible and determine the function 
$\varepsilon(a,b)$ in lowest order to be 
$\varepsilon=\varepsilon_0 -\beta_1/(\beta_2a),$   
where $\varepsilon_0$ is a scale-independent constant.  
($\beta_1$ and $\beta_2$ are constants entering the one loop beta-functions 
$\beta_\varepsilon$ and $\beta_a$, respectively, given below and in Appendix~\ref{sec:betas}.)

We should emphasize that \eqn{eq:rgi}\ and the remarks below it apply only
in the model  without matter fields.  In general, we would expect
$\beta_\varepsilon$ to be a function of  {\it all\/}  the other dimensionless
coupling constants in the theory, except, as we have described, 
$\varepsilon$ itself. Indeed, $\beta_\varepsilon$ is nonzero 
even if we do not quantize gravity, in other words there are ``pure matter'' 
contributions, independent of $a,b$.
One might expect  such
contributions to appear at two loops from graphs with two gauge
couplings  or two Yukawa couplings, and at three loops from graphs with
two quartic scalar couplings.  One sees, however, from \reference{Jack:1983sk}\
and \reference{Jack:1985wd},  that 
although such graphs generate contributions to $\beta_a$ and $\beta_b$, they do 
not contribute to $\beta_{\varepsilon}$. 

In fact, $\beta_{\varepsilon}$ as described here is 
the Euler anomaly
coefficient, that is, the coefficient of $G$ in the gravitational trace anomaly. 
It thus represents a generalization to the quantized $R^2$-gravity case of the 
candidate $a$-function proposed by Cardy\cite{Cardy:1988cwa} as manifesting a
4-dimensional $c$-theorem. Results for this anomaly coefficient
(without quantizing gravity) include  a non-zero 5-loop contribution
involving four  quartic scalar couplings\cite{Hathrell:1981zb}  and
non-zero three loop contributions involving gauge and Yukawa
couplings\cite{Freedman:1998rd}. For more on the
$a$-theorem see \cite{Jack:1990eb, Komargodski:2011vj, Komargodski:2011xv};
for some recent progress see \cite{Jack:2013sha,Jack:2014pua}, and, for some interesting 
potential cosmological consequences, see \cite{Barvinsky:2013tta}.

One consequence of our considerations is that,  even though $G$ is a
covariant divergence and $\sqrt{g}\,G$ is an ordinary derivative, it can
contribute a nonzero {\it value}\/ to the action in \eqn{eq:lho2} in
curved spacetime just like  the other terms, even though it is
equivalent to a ``surface" term or ``boundary" term. For example, in a
maximally symmetric background, $G=R^2/6.$  It is paradoxical
that a ``surface term" could be of the same order as a volume term in
the action. Even more, Euclidean de~Sitter space is topologically
the  sphere $S^4$, so that there is no boundary  or surface whatsoever,
yet the integral is nonzero, apparently violating Gauss's law.  The
resolution of this paradox is that  although $G=\nabla_\mu B^\mu$ is
gauge-invariant,  $B^\mu$ is not, \ie, it does not transform as a vector
under general coordinate transformations\footnote{In other words,
$\sqrt{g}\,G$ is closed but not exact on $S^4$. See Appendix~\ref{sec:g-b}.}.
This is related to the fact that the surface $S^4$ is 
homotopically nontrivial.

\section{Scale Symmetry Breaking and Naturalness}\label{sec:natural}

\paragraph{ }
This theory is classically scale-invariant but not conformally
invariant. The associated QFT breaks scale invariance through the
renormalization procedure by which the coupling constants become
scale-dependent. Classical scale symmetry is 
therefore anomalous in QFT; the divergence of the
dilatation current, instead of vanishing, becomes the sum of
beta-functions of couplings or masses times their corresponding
operators.  

This anomaly has nothing to do with
naturalness~\cite{'tHooft:1979bh}, which is associated with power-law
divergences, typically characterized in terms of some cutoff $\Lambda$
as quadratic behaviour $\Lambda^2$ for scalar masses or $\Lambda^4$ for
the vacuum energy, times some coupling constants. This is a physical
effect perhaps best illustrated in the context of grand unified theories
(GUTs) in which the $SU(2)\otimes U(1)$ electroweak theory is embedded
in some larger group $G$, such as $SU(5)$. The GUT theory involves
particle masses $M_U\gg M_W$, and it is difficult to arrange for the
ratio $M_W/M_U$ to be as small as required, $10^{-13}-10^{-14}$, because radiative
corrections to the lighter masses such as $M_W$ are often proportional
to the larger scale $M_U$. This provides motivation for softly broken
supersymmetry (susy), still the most popular extension of the Standard
Model (SM).  Any theory in which such effects are
suppressed seems to depend upon some symmetry to protect it.

Classically scale invariant theories, although anomalous, beget a legacy
to their corresponding QFT's. As has been emphasized by
Bardeen~\cite{Bardeen:1995kv} and others~\cite{Foot:2007iy}, the breaking of
scale invariance by anomalies is ``soft", reflecting logarithmic divergences 
of the ``bare" theory that are responsible for running couplings.
This is not true for power divergences, a radiative correction behaving,
for example, as $g^2(\Lambda)\Lambda^2/(4\pi)^2$. Even if the coupling
$g^2(\Lambda)$ were AF, it would vanish relatively slowly, as
$1/\log(\Lambda)$ as $\Lambda\to\infty,$ so that $g^2(\Lambda)\Lambda^2$
does not become small. Power-law divergences are therefore incompatible
with a theory having classical scale invariance. Turning this around,
this is why effective field theories that are intended to apply below
some physically relevant higher mass scale are not classically scale
invariant. Such models usually have radiative corrections that
behave like powers of the high scale.  
In the present circumstances, in which we wish to entertain the
possibility that there are no physically relevant higher mass scales, it
is perfectly natural to ignore potential power divergences as
manifestations of the regularization method. In fact, DREG is a
regularization procedure that does assign the value zero to power
divergences, which is the correct procedure in the present context. 

As mentioned above, not  only is the pure gravity theory AF, but it
conveys this property to the dimensionless  matter
couplings~\cite{Fradkin:1981hx,Buchbinder:1992rb} that may be added,  so
that the ultraviolet behaviour for many of these models is perfectly 
natural\footnote{A word of caution must be issued here;  in the case of
a single real field,  we find that the basin of attraction of the UV
fixed point is limited.}.

Previous workers have added an Einstein-Hilbert term $M_P^2 R$ and a
cosmological constant $\Lambda_{cc}$, thereby explicitly breaking
classical scale invariance. This theory remains formally renormalizable
and AF, since $M_P^2R$ and $\Lambda_{cc}$ are UV irrelevant operators. 
From this point of view, this looks acceptable, and, assuming
that the couplings $a,b$ are still sufficiently small on the scale
$M_P$, the effective field theory below $M_P$ will look conventional.
Flat spacetime would appear to be a sensible solution to
\eqn{eq:extrema} at large distances, but it is easily seen that
perturbations about that background have a massive spin-two field with
negative kinetic energy. This is the origin of the belief that the
theory violates unitarity. 
From another point of view, however, the addition of these irrelevant couplings to
the bare theory is a drastic
modification, since it is no longer natural to ignore power-law
divergences associated with radiative corrections. 
As a result, it would appear to require extremely fine tuning to sustain this form of the
theory, so it would be impossible to argue that it represents a UV completion
of general relativity. Consequently, this theory is unacceptable as a
starting point for a completion of gravity, and such a model must be
interpreted as an ordinary effective field theory in which the terms
quadratic or quartic in curvature are simply some of the operators
that can be expected to become important at energy scales 
on the order of $M_P$ but small compared 
to some large physical cutoff $\Lambda_{eff}$.

In order to account for ordinary Einstein gravity in a natural way, models such as the
ones considered herein, described by $S_{ho}$ plus matter, must undergo DT, 
as described in the Introduction, Section \ref{sec:intro}. 
In the next section, we review and extend the formalism for
investigating this possibility perturbatively.

\section{Dimensional Transmutation in $R^2$ Gravity}\label{sec:dtgravity}

\paragraph{ }
The formalism will be reviewed for a case that has already been
partially discussed in the 
literature~\cite{Fradkin:1981hx,Avramidi:2000pia,Buchbinder:1992rb}, 
although from a rather different perspective. For this purpose, it will be useful to
define the rescaled coupling $w\equiv a/b$, so that the action
\eqn{eq:lho2} becomes
\beq\label{eq:lho3}
S_{ho}=\int d^4x\sqrt{g}\left[ \frac{1}{a} \left(\frac{1}{2}C_{\kl\mn}^2
+ \frac{w}{3}R^2\right)+\varepsilon G\right].
\eeq
This form has several advantages. The AF coupling $a$ may also be
identified with the loop-expansion parameter, whereas the coupling $w$
will be seen to approach a UV fixed point. As we shall discuss below,
the form of the $\beta$-functions suggest treating $a$ as the primary
coupling governing the asymptotic behaviour of the others.

As usual, the investigation of spontaneous symmetry breaking (SSB) of a
theory involves the effective action  $\Gamma[g_\mn(x)]$. Like the
classical action, it is a functional of the fields. The extrema of the
effective action determine candidates for local minima, maxima, and
saddle-points:
\beq\label{eq:extrema}
\frac{\delta}{\delta g_\mn (x)}\Gamma[g_\mn]=0.
\eeq
Metrics satisfying this equation are said to be ``on-shell".
We have suppressed the dependence of $\Gamma[g_\mn]$ upon the coupling constants 
$a(\mu),w(\mu),\varepsilon(\mu)$ and the normalization scale $\mu$, but they are important. 
The effective action obeys the renormalization group equation (RGE)
\beq\label{eq:rge}
\left[\mu\frac{\partial}{\partial \mu}+\beta_a\frac{\partial}{\partial a} +
\beta_w\frac{\partial}{\partial w}+\beta_\varepsilon\frac{\partial}{\partial\varepsilon}+
\beta_{\sigma_j}\frac{\partial}{\partial\sigma_j}-
\gamma\!\! \int\!\! d^4x\,
g_\mn(x)\frac{\delta}{\delta g_\mn(x)}\right]
\Gamma[g_\mn]=0,
\eeq
where $\sigma_j$ denote possible gauge-fixing
parameters, and  $\gamma$ the anomalous dimension of the metric.
The effective action is the generator of the 1PI $n$-point functions 
$\Gamma_n(g_\mn(x_1), g_\mn(x_2),\ldots,
g_\mn(x_n))$, and it is nonlocal in general. In perturbation theory, the ``classical"
action consists of a term of the form of \eqn{eq:lho3}. Radiative
corrections consist of loop diagrams plus divergent counterterms of the
same form as \eqn{eq:lho3} such that all  $\Gamma_n[{g_\mn(x_j)}]$ remain 
finite as the cutoff is removed.  

The one-loop effective action has not been determined for an arbitrary
background metric, so it is impossible to discuss all possible solutions
of \eqn{eq:extrema}. However, it is clear that, formally, this equation
will have a solution for flat spacetime, $g_\mn=\eta_\mn$, where all
curvature tensors vanish. However, as we have remarked, we do not expect
this to be a consistent background solution of the QFT, because the
couplings become strong in the infrared, and this appears to be a
confining theory. In this respect, it is similar to Yang-Mills theory.
Just what a consistent solution looks like, we do not know, but we would
expect the emergence of a DT scale, $\Lambda_{ho}$.  Since the theory
has more than one coupling, there may remain free parameters
on which the spectrum and interactions could depend.
Whether there can be any light states below $\Lambda_{ho}$ is unclear,
but, regardless, it is highly unlikely that this theory would resemble 
conventional gravity at long distances.

Since the original theory \eqn{eq:lho3} has 
more than one coupling constant, one
may ask whether DT can occur for weak coupling. To our knowledge, this
has not been explored before. To simplify the analysis, we shall assume
that the field is maximally symmetric, so that the metric describes
either de~Sitter (dS) or anti-de~Sitter (AdS) spacetime, depending on
whether the constant curvature is positive or negative.  We shall
investigate whether DT can occur for a particular value of the curvature
$R$. For the dS case, the Euclidean manifold is usually compactified on
a  four-sphere~\cite{Avramidi:2000pia} because the isometries of dS are
the rotations $SO(5)$ (or $SO(1,4)$ for Lorentzian signature.)  The
global topology is unimportant in perturbation theory.   For Euclidean
AdS, the isometry group is $SO(1,4)$ for Euclidean signature  (or
$SO(2,3)$ for Lorentzian signature).  In this case, the associated 
spacetime is hyperbolic, so the manifold inherently has infinite volume.
In either case, the maximally-symmetric background has  $C_{\kl\mn}=0,$
$R_\mn= g_\mn R/4$. Therefore, $G=R^2/6$, and  the value of the
classical action is 
\beq
S_{ho}=\int d^4x\sqrt{g}\, \frac{R^2}{3}\left(\frac{1}{b}+\frac{\varepsilon}{2}\right).
\eeq
All we really need to know about the volume element is that $d^4x\sqrt{g}\propto 1/R^2$,
which can be inferred from dimensional analysis alone. Thus,
\beq\label{eq:sho3}
\int d^4x\sqrt{g}\equiv \frac{V_4}{R^2}, \qquad {\rm so}\ S_{ho}=V_4 \left(\frac{1}{3b}+\frac{\varepsilon}{6}\right)=\frac{V_4}{6}\left(\frac{2w}{a}+\varepsilon \right),
\eeq
where $V_4$ is some dimensionless volume element independent of $R.$ 
For the four-sphere of dS, $V_4=6(8\pi)^2,$
while for (the  cover of) hyperbolic AdS, it is infinite,
so we have to imagine a temporary large distance cutoff so that the
spacetime has a finite volume. Alternatively, we can isolate the reduced
effective action $S_{ho}/V_4$, which is the analog of the
effective potential in flat spacetime.

Consider the calculation of the effective action $\Gamma[g_\mn^{B}]$ by the
background field meth\-od\footnote{The background field method is 
reviewed briefly in Appendix~\ref{sec:bfm}.},
which involves shifting the metric
$g_\mn=g_\mn^{B}+h_\mn$ by a classical field and treating $h_\mn$ as
the quantum field\footnote{Alternate definitions of the quantum field
are sometimes used. See Ref.~\cite{Buchbinder:1992rb} for further discussion.}. The
background field in our case will be assumed to have maximal symmetry,
but the quantum field over which we integrate is arbitrary. 
In order that fluctuations about
the background be stable, there may be restrictions on the 
coupling constants. For example, 
Avramidi~\cite{Avramidi:2000pia} showed that the couplings
must obey the constraints, \eg, $a(\mu)>0,$ 
$0<w(\mu)<3/2$, for convergence of the Euclidean path integral, 
but he did not indicate at what scale $\mu$ such inequalities must hold. 
We shall return to these issues in Section~\ref{sec:constraints}; see also
Appendix~\ref{sec:stability} for further details.

For a maximally symmetric background, the only unknown quantity is the
magnitude of the curvature $R$. In this paper, we shall only 
investigate in detail the case of positive curvature, leaving AdS for
later work. Let us call $\rho\equiv\sqrt{R\,}.$ The question is whether
$\rho$ may be determined by DT. The effective action $\Gamma$ 
can depend only on $\rho,\mu,a(\mu),w(\mu),\varepsilon(\mu).$   In fact,
given \eqn{eq:varindex}, only the ``classical" action  can depend on the
parameter $\varepsilon$.  Therefore, it will not enter the Feynman rules
for calculating radiative corrections $\Delta\Gamma$ to the effective
action.  Nevertheless,   $\varepsilon$ is renormalized and does require
counterterms which, however, only depend on the other coupling
constants. (For further discussion, see  Appendix~\ref{sec:bfm}.)

Since $\Gamma$ is dimensionless, its scale dependence must
be in terms of the ratio $\rho/\mu$.  We may therefore express its  
loop expansion in the following form\footnote{In general, there will also
be a term $A(a,w)$ on the right-hand side representing finite local
counterterms characteristic of the particular renormalization scheme.
Even for minimal subtraction (MS), it is nonzero. We shall assume that
the renormalization prescription has been modified in such a way as to
remove such terms, which, while they could be included, only serve to
complicate our subsequent discussion. See below, however, 
concerning the possibility of an imaginary part of $\Gamma.$ }: 
\beq\label{eq:gloops} 
\Gamma(\rho)\!=\!S_{ho}(a,w,\varepsilon)\!+\!B(a,\!w)\log(\rho/\mu)\!
+\!\frac{C(a,\!w)}{2}\log^2(\rho/\mu)\!+\!\frac{D(a,\!w)}{6}\log^3(\rho/\mu)+\ldots.
\eeq
The coefficients $B, C, \ldots$ are functions of the dimensionless
couplings $(a(\mu),w(\mu))$, but the dependence on $\log\mu$ has been
exhibited explicitly. In the loop expansion, 
\beq\label{eq:loops}
B(a,w)\!\equiv\!\! \sum_1^\infty B_k(w)a^{k-1}\!,\ C(a,w)\!\equiv\!\! \sum_2^\infty C_k(w)a^{k-1}\!,\ D(a,w)\!\equiv\!\! \sum_3^\infty D_k(w)a^{k-1}\!, \ldots,
\eeq
where the coefficients $B_k$, $C_k$, $D_k$, etc., represent the
contribution $k$-th order. In general, the coefficient of the power 
$\log^n(\rho/\mu)$ is nonzero beginning at loop-order $k=n$. At
one-loop, only the term with coefficient $B$ arises; at two-loops, the
term having coefficient $C$ also arises, etc. The first derivative
of the effective action is 
\beq\label{eq:gamma1}
\frac{\partial \Gamma}{\partial\rho}
=\frac{1}{\rho}\left[B(a,w)+C(a,w)\log(\rho/\mu)+\frac{D}{2}\log^2(\rho/\mu)+\ldots\right].
\eeq
An extremum at $\rho=v\ne0$ satisfies \eqn{eq:extrema}, which in the
present application, reduces to an ordinary derivative, $\Gamma'(v)=0$
in \eqn{eq:gamma1}. Obviously, this equation simplifies considerably if
we choose to normalize at $\mu=v$:
\beq\label{eq:Gmin}
\frac{\partial \Gamma}{\partial\rho}\Big|_{\rho=v}=\frac{1}{v}\left[B(a(v),w(v))\right]=0.
\eeq
The meaning of this stark equation is that, given the function $B(a,w)$,
an extremum will occur if one can find a scale $v$ at which the
couplings are related according to the equation $B(a(v),w(v))=0.$ At
one-loop order, this corresponds to a value of the coupling $w=w_1$,
where $B_1(w_1(v))=0,$ independent of $a$! 

To characterize this extremum as a local maximum or minimum, we must know
\beq\label{eq:curvature}
\delta^{(2)}\Gamma=\half\Gamma''(v)(\delta\rho)^2=\frac{1}{2v^2}C(a(v),w(v))(\delta\rho)^2,
\eeq
where $C$, we recall from \eqn{eq:loops}, starts at two-loop order
$C_2(w)a.$  This is in fact the mass of the dilaton which arises from the scale-breaking anomaly.
In fact, we shall see that $C_2$ can be determined from
one-loop results. Because we have assumed such a simple background, one
can determine the form\footnote{In ref.~\cite{Avramidi:2000pia}, Avramidi calculated
$B_1$ by explicitly evaluating the functional determinants arising at one-loop. See Appendix~\ref{sec:bfm} for further discussion. He checked
his result by showing that it satisfied the RGE. Our result for $C_2$ is new.} of $B_1(w)$ and
$C_2(w)$ directly using the RGE, \eqn{eq:rge}, which we write in the form
\beq\label{eq:rge2}
-\left[\mu\frac{ \partial}{\partial \mu}-\gamma\rho\frac{\partial}{\partial\rho}\right] \Gamma(\rho)
=\left[\beta_a\frac{\partial}{\partial a} +
\beta_w\frac{\partial}{\partial w}+\beta_\varepsilon\frac{\partial}{\partial\varepsilon}\right]
 \Gamma(\rho)+\ldots.
\eeq
Using the fact that $\Gamma(\rho)$ depends on $\rho$ only through the
ratio $\rho/\mu$, the left-hand side may be written as 
$(1+\gamma)\rho\,\partial\,\Gamma/\partial\rho$. 
The only dependence on $\varepsilon$ is through the
``classical'' action, \eqn{eq:lho3}, (including counterterms), and 
$\beta_\varepsilon$ is related to the other beta-functions through \eqn{eq:rgi}.

We have suppressed the gauge-dependent terms on the right-hand side of \eqn{eq:rge2}, as
they will not affect our results. 
We will find that the RG equation relates $B(a,w)$, $C(a,w)$ etc to the $\beta$-functions 
{\it and\/} the 
{\it \/} gauge-dependent anomalous dimension $\gamma$. However 
the dependence on $\gamma$ cancels out in {\it on-shell\/} (i.e. physical) quantities.
In the case involving a matter field, to which we will turn in the next 
section, this cancellation is quite nontrivial because (as we shall see) in that case 
both $B_1$ and $C_2$ depend on $\gamma$ in general. 

It was observed long ago that the RGE relates different orders
of the loop-expansion for $\Gamma$~\cite{'tHooft:1973mm}. 
The beta-functions and anomalous dimensions have loop expansions of the
same form as $B(a,w)$ in \eqn{eq:loops}, so, if they are known to some
order, then one may insert them into the RGE \eqn{eq:rge}, together with
the loop expansion in \eqn{eq:gloops}, and equate common powers in
$a^n$ (or $\hbar^n$).  
Thus (inserting explicit factors of $\hbar$ for clarity) we find 
\begin{align}
\label{eq:b1rg}
\mu\frac{ \partial\Gamma }{\partial \mu} &= -\hbar B_1 -\hbar^2 B_2 
- \hbar^2 C_2\ln (\rho/\mu)+\cdots\\
\sum_i\beta_i\frac{\partial\Gamma}{\partial \lambda_i}
&=  \sum_i\left(\hbar\beta^{(1)}_i+\hbar^2\beta^{(2)}\right)\frac{\partial}{\partial \lambda_i}S_{ho}
+ \hbar^2 \sum_i\beta^{(1)}_i\frac{\partial}{\partial \lambda_i}B_1\ln (\rho/\mu) +\cdots\\
-\gamma\rho\frac{\partial}{\partial \rho}\Gamma &= -\hbar^2\gamma^{(1)}B_1 + \cdots 
\end{align}
where $\beta^{(1)}_i$ denotes the one-loop beta-function for the 
coupling $\lambda_i$, $S_{ho}$ is given in \eqn{eq:sho3},  and the sums
are over all couplings on which the classical action depends $\{a,w,\varepsilon\}.$ 
It follows that 
\begin{align}
\label{eq:b1ho}
B_1&=\sum_i\beta^{(1)}_i\frac{\partial}{\partial \lambda_i} S_{ho},\\
\label{eq:c2ho}
aC_2&= \sum_i\beta^{(1)}_i\frac{\partial}{\partial \lambda_i}B_1=
\left[\sum_i\beta^{(1)}_i\frac{\partial}{\partial \lambda_i}\right]^2 S_{ho},\\
\label{eq:gaho}
aB_2 &=\beta^{(2)} \frac{\partial}{\partial \lambda_i}S_{ho}-\gamma^{(1)}B_1.
\end{align}
Thus, from \eqns{eq:b1ho}{eq:c2ho}, we obtain the leading contributions to both 
the condition for an extremum (\eqn{eq:Gmin}) and its nature (\eqn{eq:curvature})
\footnote{One can check that the results in
\eqns{eq:b1ho}{eq:c2ho} are unchanged by the addition of finite local
counterterms $A_0(a,w)$. That will not be true for the two-loop
contributions to $B$, for example.}. Note that neither condition depends on 
the anomalous dimension $\gamma$.

In MS, in each order of
the loop expansion, the only really new contribution is to the single
log term, $B$, with all the higher powers of $\log(\mu)$ determined by
lower-order corrections\footnote{One can check that the results in
\eqns{eq:b1ho}{eq:c2ho} are unchanged by the addition of finite local
counterterms $A_0(a,w)$. That will not be true for the two-loop
contributions to $B$, for example.}.

There is a possible flaw in the preceding method of determining the
effective action.  Although, as mentioned in an earlier footnote, it is
possible to adopt a renormalization prescription to remove real local
counterterms $A(a,w),$ if the effective action had an imaginary part of
this form, it could not removed by counterterms.   If present, an
imaginary part must be regarded as an instability.   In a direct
evaluation of the functional determinants, it would show up as a
negative eigenvalue that would prevent one from carrying out the path
integral.   Just as in flat space models, such as scalar $\lambda\phi^4$
theory  with a negative $m^2\phi^2$ term, such a term could arise by
continuation of the effective potential from a region where there is no
imaginary part to another region where the argument of a logarithm turns
negative\footnote{Avramidi~\cite{ Avramidi:2000pia} actually did
evaluate the functional determinants for a similar model that included
an Einstein-Hilbert term as well as a cosmological constant.  We can
take advantage of his calculation to check some of our results, but he
did not evaluate the curvature $C_2$.  One can in principle obtain the
results below from his by forming the RG-improved effective action
starting from his one-loop effective action.}.   Another potential
shortcoming of this method is that it does not reveal whether there are
zero modes.  In fact, as shown in~\cite{Avramidi:2000pia}, there are
five in the conformal sector of the metric fluctuations.

Of course, once one has a formula for the one-loop corrections to the
real part of the effective action via the RGE, one can check whether or
not fluctuations are unstable and whether there remain flat directions,
and this can be done without performing any functional integrations.  We
shall discuss this further in Section~\ref{sec:constraints}.

To apply these formulas, we need the one-loop beta-functions
~\cite{Avramidi:1985ki}; 
\begin{subequations}
\label{eq:betaho}
\begin{align}
\label{eq:betahoae}
\frac{1}{\kappa}\beta_a^{ho}&=- \beta_2^{ho} a^2,\ 
\beta_2^{ho}=\frac{133}{10},\qquad 
\frac{1}{\kappa}\beta_\varepsilon^{ho}=-\beta_1^{ho},\ 
\beta_1^{ho}=\frac{196}{45},\\
\label{eq:betahow}
\frac{1}{\kappa}\beta_w^{ho}&=\frac{10\,a}{3}\left[w^2-\frac{549}{100}\,w+\frac{1}{8} \right],
\end{align}
\end{subequations}
where $1/\kappa\equiv16\pi^2.$ For $a>0,$ $\beta_a$ displays AF, as claimed, and
the sign of $a$ is renormalization group (RG) invariant . In order to
have a Euclidean action bounded from below, we require $a>0$.  The
running of $w$ is more complicated; $\beta_w$ has two real zeros. 
There is a UV fixed point at $w_1\approx 0.023,$ $(a\ll
b)$, and an IR fixed point at $w_2\approx 5.47$ $(a>b).$  
Naively, it appears as if 
$w=0$ is neither a singular point nor a
fixed point, but $1/w=b/a\to\infty$ as $w(\mu)\to0$, and therefore
$b\to\infty$ (a ``Landau" pole). 
Since perturbative corrections are polynomials in the parameters
$(a,b)$, this constitutes a breakdown of perturbation theory.  
Typically, we expect perturbation theory to be valid only for
$\kappa a\ll1$ and $\kappa b\ll1,$ so we cannot trust the one-loop 
results arbitrarily near $w=0.$

On the other hand, $w\to\infty$ corresponds to $b\to0,$ which is {\it not} a 
breakdown of perturbation theory.  
It would have been better to take the ratio of couplings as
$\widetilde{w}=1/w=b/a$, since perturbation theory  holds as
$\widetilde{w}\to0,$ but we shall continue to follow past conventions.

To determine possible extrema perturbatively, we may evaluate
\eqns{eq:b1ho}{eq:c2ho} in the one-loop approximation, yielding:
\begin{align}
\label{eq2:b1ho}
B_1&=V_4\frac{10\kappa}{9}\left[w^2-\frac{3}{2}w-\frac{317}{600}\right],\\
\label{eq2:c2ho}
C_2&=V_4\beta_w\frac{20\kappa}{9}\left(\!w-\frac{3}{4}\right)
=V_4\,a\frac{200\kappa^2}{27}\left[w^2-\frac{549}{100}\,w+\frac{1}{8} \right]
\!\left(\!w-\frac{3}{4}\right).
\end{align}
The extrema occur where $B_1=0$, \viz, $w_\pm=3/4\pm\sqrt{3927}/60$. 
For both the positive root, $w_+\approx 1.794$ and the 
negative root $w_-\approx -0.294$, we find from \eqn{eq2:c2ho} 
that $C_2 < 0$, and hence (from \eqn{eq:curvature}) that both 
extrema are local maxima of the action. 

Even though they are not locally stable, we would like to determine whether these extrema can be reached naturally in the course of the running of coupling constants or whether fine-tuning would be required to arrange for these values of the coupling constants.  This may be less interesting than if they were minima, but the analysis serves to illustrate concepts useful in models having additional coupling constants with more complicated renormalization flows.  Further, as we shall discuss, it is conceivable that maxima such as these and saddle points could be cosmologically relevant.

To discuss the running of the couplings, we shall assume that the initial values 
$a_0, b_0$ are sufficiently small so that perturbation theory may be used 
at the starting point.  In view of the fixed points at $w=w_1$ and $w=w_2$, there are
three possible phases to be discussed: 
(1)~$w_1<w(\mu)<w_2,$ 
(2)~$w(\mu)>w_2$  or $w(\mu)<0,$ and
(3)~$0<w(\mu)<w_1.$    
  As noted, we would expect perturbation  theory to be valid so long as 
$\kappa a(\mu)\ll1$ and $\kappa b(\mu)\ll1,$ and any initial value of the ratio 
$w_0=a_0/b_0$ can be accommodated perturbatively except for $w\to0$, 
where $b\to \infty$. 
\begin{enumerate} 
\item
Starting at any value $w_0$ between the two fixed points, $w_1<w_0<w_2$,
$w(\mu)$ spans the entire region by running toward higher or lower
scales $\mu$.   The extremum 
corresponding to $w = w_+\approx 1.79$ lies within this region 
and will be accessible perturbatively at some scale $v$, so long as $\kappa a(v)\ll 1.$
Perturbation  theory certainly holds in a neighborhood of $w_1\approx 0.023$, even though $b\gg a,$ since both $a(\mu)$ and $b(\mu)$ vanish as $\mu\to\infty.$  
Perturbation theory will break
down as $\mu$ decreases, since $a(\mu)$ monotonically increases.  Although formally $w(\mu)\to w_2$ as $\mu\to0,$
the theory will become strongly coupled at some finite value of $\mu$. 
\item 
Starting at any $w_0$ for $w_0>w_2$, we see that $w(\mu)\to+\infty$ as the scale $\mu$ increases.  Since $a(\mu)$, $b(\mu)$ are decreasing, this does {\it not}\/ constitute a breakdown of
perturbation theory, but it simply means that $b(\mu)$ passes through zero at some finite value of 
$\mu.$  Since $b=0$ is not a fixed point, as $\mu$ increases further, $b(\mu)$ turns negative, and therefore also $w(\mu)<0$. $w(\mu)$ continues increasing through negative values toward the extremum at $w=w_-\approx-0.29.$  So long as $|b|$ does not become too large, this could remain within the reach of perturbation theory.  A similar story obviously holds if the starting value is in the region $w_0<w_-.$  The couplings are continuous at $w=\infty,$ so this point should be thought of as compactified. 

If the initial value $w_-<w_0<0$, then one must decrease the scale $\mu$ to run toward $w_-.$  Whether the DT scale $v$ can be reached will depend on whether perturbation theory continues to hold as $a$ increases and $|b|$ decreases.
\item 
With $0<w_0<w_1\approx+.023,$ there is no extremum of the action in this region, so the behavior of the couplings is irrelevant for DT in this ``pure gravity" model.  Nevertheless, for completeness, we shall remark on the running.  As the scale increases, $w\to w_1,$ and $a$ and $b$ are AF.  
Decreasing the scale runs toward the scale where $b\to+\infty,$ and perturbation theory breaks down.   We  guess this would occur for $\kappa b(\mu)\sim 1.$ To be slightly more quantitative,
if the initial value $\kappa b_0\sim 1$, then $\kappa a_0\sim w_0<w_1$, or
$a_0<w_1/\kappa\approx 3.61.$  The range of validity of perturbation
theory therefore depends on how much smaller $a_0$ is than this.   
\end{enumerate}

In summary, we have found that there are two extrema at scales $\mu=v$
determined, for $b(v)>0$ by $a(v)/b(v)=w_+\approx1.79$ and another, for
$b(v)<0$, by $a(v)/b(v)=w_-\approx -0.294$.  Both can be reached in
perturbation theory starting from a wide range of initial values;
however, both are local maxima since $C_2<0$, \ie, the dilaton is
tachyonic.   By our method of calculation, we cannot tell whether these
extrema occur for $R>0$ (de~Sitter-like) or $R<0$ (anti-de~Sitter-like),
but there is good reason to  presume that it is valid in de~Sitter
background.  

Since these are metastable vacua in de~Sitter background, they might be
candidates for ``new inflation" scenarios if the local maxima are
sufficiently flat. A quantitative measure of the degree of flatness in
conventional models is the slow-roll parameter $\eta=M_P^2
V''(\phi)/V(\phi),$ where $V(\phi)$ is the potential at the field value
of interest.   By transforming to Einstein frame, one can show that the
corresponding quantity in our model is $\eta=m_d^2/\Lambda,$ where $m_d$
is the dilaton mass proportional to $C_2,$ and $\Lambda$ is the
corresponding cosmological constant.  Having determined $R=v^2$ from
$B_1(w)=0,$ we find that $\eta=2C_2 v^2/(3M_P^2),$ where $C_2$ is given
in \eqn{eq2:c2ho}. (The appearance of $M_P$ here is due to the fact 
that in the Einstein frame the theory takes the form of Einstein gravity with 
a positive cosmological constant, coupled to a massless scalar.) 
As discussed earlier, the dilaton mass arises from
the scale anomaly at two-loop order, so with reasonable values of the
couplings,  one can expect $\eta$ to be small. We have not analyzed this
model at finite temperature, and we do not know the limit on $a(v)$ that
would allow sufficient inflation
\footnote{We note here that there has been  recent work on inflation in
the $R^2$ gravity  context, in the light of the recent BICEP2
data\cite{Rinaldi:2014gua},\cite{Rinaldi:2014gha}.
For other recent work on $R^2$ gravity and its supersymmetric extensions, see
\reference{Kounnas:2014gda}}.

Although this is not a realistic model of our universe, this is a rather 
different inflationary mechanism than has been encountered previously. The metric
is in a sense self-inflating. Of course, unlike Einstein-Hilbert theory,
the metric in this model has additional degrees of freedom beyond the
massless graviton, including a scalar mode, but it is not obvious that that this
mode may be identified as the inflaton. Nevertheless, without any fine
tuning, this already has some of the ingredients of a successful
inflationary model, except, of course, that it is unlikely to exit to a
phase that resembles general relativity, a problem that may be cured with the
introduction of matter.
 
\section{Matter: The Real Scalar Field}\label{sec:realscalar}
 
 \paragraph{ }
In order to obtain a realistic field theory of gravity, it seems
necessary to include matter fields. We shall simply discuss a real
scalar field here, leaving the addition of other scalars, gauge fields
and fermions for later work. The hope is that the matter action
\eqn{eq:jrealscalar} will lead to a nonzero \vev\ for $\phi$, so that we
may identify $\xi\phi^2$ with the reduced Planck scale $M_P^2/8\pi$.  
The idea of generating the Planck mass in this way is not original to
us; indeed, in the final section of the ref.~\cite{Fradkin:1981hx},
those authors suggested that it would be interesting to explore these
possibilities and, in a footnote, provided the formula for the one-loop
correction. The idea is to have a CW-like model with gravity replacing
electrodynamics in its effect on the scalar field.  This idea was
followed up in a number of  papers~\cite{Odintsov:1989gz,
Elizalde:1994gv,Elizalde:1995at}; however, there was never completed a
fully self-consistent calculation that included quantum corrections to
the background metric.  Often feedback on the metric was assumed to be
negligible. 

Our approach is fundamentally different from previous treatments in
several respects. We insist that the starting theory be classically
scale invariant, so that this theory of gravity can be entertained as
potentially complete without naturalness problems.  The background curvature is to
be determined self-consistently\footnote{A similar scheme was 
attempted in Einstein gravity in \cite{Fradkin:1983mq}.}, and the Planck mass must be generated
dynamically via DT. One reason for first discussing the gravitational
theory without matter in the previous section was to gain some
experience with the gravitational dynamics of such a model before embarking on other
scenarios. So far, we have only treated maximally symmetric models,
but, in principle, more complicated gravitational backgrounds can be
considered. Our failure to find a locally stable vacuum state in the
preceding section is consistent with the view that, without matter,
gravity is bootless. Perhaps they can be tied together in a grand
unified framework, but we feel there is much to be learned first in
simpler models of this type before attempting that.

If DT does occur, then below the Planck scale, the theory will take the
form of an effective field theory resembling the usual sort of
scalar-tensor theory of gravity but with calculable corrections or
matching conditions specified. We anticipate that the naturalness
issues associated with physics below the Planck scale would return, so
we cannot immediately suggest that this approach is a solution to the
naturalness problems of particle physics. (Of course, as with supersymmetry
breaking, 
one could arrange for this dynamics to be in a sector hidden from the 
Standard Model~\cite{Foot:2007iy,Foot:2010av}.)

The action we shall consider is the sum of \eqn{eq:jrealscalar} and
\eqn{eq:lho2}, $S=S_{ho}+S_m$. This classical action has no masses and
is formally invariant under global scale transformations, which we
define as 
\beq\label{eq:scaletransf}
\phi(x)\to e^{\alpha}\phi(x),\quad 
{g}_\mn(x)\to e^{-2\alpha}{g}_\mn(x),
\eeq
as reviewed in Appendix~\ref{sec:scaleinv}.

The EoM associated with this action are 
\begin{subequations}
\begin{align}
\begin{split}
\hskip-15mm -\left(\frac{2}{3b} R - \frac{\xi\phi^2}{2}\right)\!R_\mn
&+\frac{{g}_\mn}{2}\left(\frac{1}{3b} R-\frac{\xi\phi^2}{2}\right)\!R \cr
&\hskip-15mm +\frac{1}{a}\left[\frac{2}{3}RR_\mn 
- 2 R^\kl R_{\mu\kappa\nu\lambda}+\frac{ {g}_\mn}{2}\left(R_\kl^2
-\frac{1}{3}R^2 \right)\right]=
\end{split}
\label{eq:eom1A}\\
\begin{split}
\half {T}_\mn\!-\!\left({\nabla}_\mu{\nabla}_\nu
-{g}_\mn\Box\right)\! &\left(\frac{2}{3b} R 
-\!\frac{\xi\phi^2}{2} \right)\!-
\frac{1}{6a}\left( 2{\nabla}_\mu{\nabla}_\nu R+\!
{g}_\mn\Box R-6\Box R_\mn \right) \!, \cr
{\rm where}\ {T}_\mn \equiv {\nabla}_\mu\phi {\nabla}_\nu\phi
&-{g}_\mn\left[\frac{1}{2} ({\nabla}\phi)^2 
+\frac{\lambda}{4}\phi^4 \right],
\end{split}
\label{eq:tmnA}\\
&\hskip-45mm{\rm and}\nonumber\\
-\xi\phi R-\Box\phi+\lambda\phi^3&=0.\label{eq:eom2A}
\end{align}
\end{subequations}
If we take the trace of \eqns{eq:eom1A}{eq:tmnA} and 
combine the results, we find
\footnote{The terms in $a$ drop out because of classical conformal invariance.}
\begin{align}
-\xi\phi^2 R+\left({\nabla}\phi\right)^2+\lambda\phi^4=
\Box\left(\frac{4}{b} R-3\xi\phi^2\right).
\end{align}
Writing $\Box\phi^2=2\phi\Box\phi+2(\nabla\phi)^2,$ we may rearrange the preceding equation as 
\begin{align}
-\xi\phi^2 R+6\xi\phi\,\Box\phi+\lambda\phi^4=
\frac{4}{b}\Box R-\left(1+6\xi\right)(\nabla\phi)^2.
\label{eq:eom3b}
\end{align}
Only for the conformal values, $(1/b)\to0,$  $6\xi=-1$, is \eqn{eq:eom3b} equal to $\phi$ times 
the scalar EoM, \eqn{eq:eom2A}, for arbitrary $\phi$. However, there are other solutions of these two equations that are mutually compatible. For example, if $\phi=\phi_0,$ a constant value, then the gradients vanish, so \eqn{eq:eom2A} implies that 
$\xi\phi_0 R_0=\lambda\phi_0^3.$ Assuming that $\phi_0\ne0,$ the scalar curvature takes the constant value $R_0=\lambda\phi_0^2/\xi.$ 
Then the trace equation \eqn{eq:eom3b} is also satisfied.
Returning to the tensor EoM, \eqn{eq:eom1A}, for constant $\phi_0, R_0,$ one can show after considerable algebra that this equation is also satisfied and yields no further information. 
In sum, constant $\phi_0$ with $R_0\!=\!\lambda\phi_0^2/\xi$ satisfies the classical EoM.  
This is a flat direction in the space of fields.  As with the model without matter, 
the value of $R_0$ is classically undetermined, since the model remains scale invariant.

Off-shell, for arbitrary constant $\phi$ and $R$, the matter action \eqn{eq:jrealscalar} and its derivatives take the form
\beq\label{eq:offshellreal}
\frac{S_m(r)}{V_4}=\frac{1}{R^2}\left[\frac{\lambda\phi^4}{4}-\frac{\xi\phi^2}{2}R\right]=
\frac{1}{4}\left[ \lambda r^2-2\xi r \right],\ S_m'(r)=\frac{1}{2}\left[\lambda r-\xi\right],\ S_m''(r)=\frac{\lambda}{2},
\eeq
where $r\equiv\phi^2/R$. This is the matter action and its derivatives for arbitrary ratio $r$, \ie, off-shell. It has an extremum for $r=\xi/\lambda,$ which is a local minimum only if $\lambda>0.$ 
(Subsequently, while searching for extrema of the effective action,
we must keep in mind that $\lambda \ge 0$ for a classically stable
ratio.) Adding the value of the gravitational action \eqn{eq:sho3}, the
total action takes the on-shell value
\beq\label{eq:sonshell}
S/V_4=\frac{1}{6}\left[\frac{2w}{a}+\varepsilon -\frac{3\xi^2}{2\lambda} \right].
\eeq
Since the scale of the fields is undetermined at the tree level, the
value of $\mu$ at which we are to evaluate the coupling constants is
unknown.  We can hope that both of these issues will be resolved by
calculating the one-loop correction to the effective action and looking
for a consistent DT solution.  Since we now have dependence on
$\{\lambda,\xi\}$ as well as the  pure gravity couplings,  the
possibilities are much richer than in the model without matter.

Before embarking upon a fairly lengthy discussion and calculation, it
may be useful to describe where we are headed. Assuming that the
field $\phi\ne0,$ we can restore minimal coupling of the scalar field by
transforming the matter action from the Jordan form,
\eqn{eq:jrealscalar}, to the so-called Einstein frame by means of a
conformal transformation  $g_\mn(x)\to \Omega^{-2}g_\mn(x)$, where
$\Omega^2\equiv \phi^2/M^2,$ where $M$ is an arbitrary unit of mass
introduced to keep the metric dimensionless. Then, after making this
substitution, the classical matter action becomes 
\beq\label{eq:erealscalar}
S_m^{\,\ecal}=\int d^4x\sqrt{g}\left[ \half (\nabla\zeta)^2
+\frac{\lambda M^4}{4}-\frac{\xi M^2}{2}R \right],
\eeq
where we defined $\zeta\equiv M\sqrt{6\xi+1}\,\log(|\phi|/M)$. We
recognize the linear term in $R$ as the Einstein-Hilbert action for
gravity with $1/ G_N=8\pi\xi M^2\equiv M_P^2$. Remarkably, the conformal
transformation has transmogrified the self-interaction of the scalar
into a cosmological constant. For $\lambda(\mu)>0$, such a model would
naturally produce inflation\footnote{As remarked below \eqn{eq:offshellreal},  
we must have $\lambda(v)>0$ for classical stability.}. 
In units of the
Planck mass $M_P$, the cosmological term is $\lambda
M^4/4=(\lambda/\xi^2) [M_P^2/(16\pi) ]^2.$ 

The original massless scalar has morphed into a massless dilaton
$\zeta\propto\log|\phi|,$ whose presence can be easily understood. The
assumption that $\phi\ne0$ corresponds to spontaneous breaking of scale
invariance, and, since scaling is a valid symmetry classically, we must
get a Goldstone boson in the broken phase. On the other hand, since the
scale symmetry is explicitly broken by the anomaly in the QFT, we would
expect the dilaton will actually have a nonzero mass that will be
parametrically small compared to $M_P$. (In fact, this mass will be shown to arise
at two loops.)

As for the terms quadratic in curvature, in Einstein frame they become 
\begin{subequations}
\begin{align}\label{eq:hoeinstein2}
S_{ho}^\ecal&=\int d^4x\frac{\sqrt{g}}{a} \left[ \frac{1}{2}C_{\kl\mn}^2+ 
\frac{w}{3} \widetilde{R}^2+a\varepsilon\widetilde{G}\right] , {\rm where} \\
&\hskip10mm\widetilde{R}\equiv R - \frac{6}{M\sqrt{\left(1+6\xi\right)} }\,\Box\zeta
+\frac{6}{M^2\left(1+6\xi \right)} \left(\nabla\zeta\right)^2,\\
& \hskip10mm \widetilde{G} \equiv G-8\nabla_\mu J^\mu,
\end{align}
\end{subequations}
the term involving the Weyl tensor being invariant under conformal
transformations. Assuming the conformal transformation does not
change the Euler characteristic of the background topology, the change
in $G$ must be of the form of a covariant  derivative of a
vector\cite{Fradkin:1985am}.\footnote{We found  $J_\mu=
\vartheta_\nu \nabla_\mu \vartheta^\nu -
\vartheta_\mu(\nabla\!\cdot\!\vartheta)+(R_\mn-g_\mn R/2)\vartheta^\nu +
\vartheta_\mu \vartheta^2,$ where
$\vartheta_\mu\equiv\nabla_\mu\log(\Omega)$, differing slightly from the result of 
\reference{Fradkin:1985am}.  The change $G \to \widetilde{G}$ plays no role 
in perturbation theory.} At energies below the Planck scale, $\sqrt{\xi}\,M,$ this takes the form of higher 
derivative terms in an
effective field theory dominated by $S_m^\ecal$, \eqn{eq:erealscalar}. For
energies of order $M_P,$ the situation becomes more subtle. As usual,
in a flat background, it would appear as if there is a graviton plus
massive scalar plus a massive spin-two ghost. However, because of the
cosmological constant, $\lambda M^4$, Minkowski space is not a solution
of the field equations, so the flat space interpretation may not be
relevant. On the other hand, this depends on the size of the
cosmological constant in units of the Planck mass, of order
$\lambda/\xi^2$. If this ratio were small, as subsequent calculations
suggest it might be, then it does seem as if there is a range of
momenta, $(\lambda/\xi^2)^{1/4}<p/M_P<1,$ where the background curvature
might be negligible. However, the ghost mass is at the upper limit of
the range of applicability of this analysis, so it is not so clear that
the implied violation of unitarity is physically observable, even in
principle. This regime is also subject to Hawking radiation from the
horizon, which may cloud the issue further, although the temperature is
relatively small. We are left uncertain but concerned about unitarity on
the Planck scale. As a final comment concerning the Einstein frame, we note 
some recent work~\cite{Kamenshchik:2014waa} concluding that (at least at one loop), results in the two frames 
(Jordan and Einstein) coincide {\it on-shell}.

To determine whether DT takes place, it is easiest to work with the action in Jordan form \eqn{eq:jrealscalar}, to which we return. As before, we write the effective action as 
\beq\label{eq:philoops}
\Gamma(\lambda_i,r,\rho/\mu) = S(\lambda_i,r) + B(\lambda_i,r)\log(\rho/\mu) + \frac{C(\lambda_i,r)}{2}\log^2(\rho/\mu) +\ldots,
\eeq
where, again, $\rho=\sqrt{R}.$  The collection of dimensionless coupling constants $\{a,w,\varepsilon, \xi,\lambda\}$ has been denoted by $\lambda_i$.  With these conventions, the value of the effective action for $\rho=\mu$ is simply the classical action, $\Gamma(\lambda_i,r,1)=S(\lambda_i,r)\equiv S_m(\lambda_i,r)
+ S_{ho} (\lambda_i)$.

As mentioned earlier, although the RGE does provide an easy way to
determine the coefficients $B(\lambda_i,r)$ and $C(\lambda_i,r),$ it is
not really a substitute for the path integral calculation.  In
particular, the effective potential might have an imaginary part that
cannot be removed by finite local counterterms but would not show up by
this method. In fact,  such contributions have plagued previous attempts
to include matter~\cite{Odintsov:1989gz}, in
particular, because of mixing with the conformal mode of the metric. In
Appendix~\ref{sec:stability}, we have checked by explicit calculation
that there are no such modes arising in this model, so the following
calculation does indeed give the correct result.  However, we also learn
that to avoid unstable modes (negative eigenvalues,) we must have all
couplings  $a, y, \xi, w$ positive and $0<w<3/2+3\xi^2/(4y),$ at least
for some range of renormalization  scales $\mu$.  We again find five
zero eigenvalues associated with $h_1$ in the conformal sector.  Must
these inequalities prevail at the DT scale?  We shall return to this
issue in Section~\ref{sec:constraints}.

It is slightly simpler algebraically to express the variations of the action in terms of $r$ and $\log\rho$ rather than in terms of $\phi$ and $R$. Since we seek solutions for which $(r,\rho)=(r_0,v)\ne0,$ there is no loss of generality in so doing. The first derivatives are
\begin{subequations}
\begin{align}
\label{eq:eomr1}
&\frac{\partial}{\partial r}\Gamma(\lambda_i, r, \rho/ \mu)\!=\!
\frac{\partial}{\partial r}S_m(\lambda_i,r)
\!+\!\log\!\left( \rho/ \mu\right)
\frac{\partial}{\partial r}B(\lambda_i,r)\!+\!
\frac{\log^2\!\left( \rho/ \mu\right)}{2} \frac{\partial}{\partial r}C(\lambda_i,r)+\!...,
\\
\label{eq:eom1}
&\rho\frac{\partial}{\partial\rho}\Gamma(\lambda_i, r, \rho/\mu)=B(\lambda_i,r)
+C(\lambda_i,r)\log\left( \rho/ \mu\right)+\ldots .
\end{align}
\end{subequations}
Note that ${\partial}S_m(\lambda_i,r)/{\partial r}$ is identical
to $S_m'(r)$ in \eqn{eq:offshellreal}.
Setting $(r,\rho)=(r_0,v)$ where these both vanish, and choosing the normalization scale $\mu=v$,
we find
\begin{subequations}
\label{eq:eomonshell}
\begin{align}
\label{eq:eomronshell}
\frac{\partial}{\partial r}\Gamma(\lambda_i, r, \rho/ \mu)\Big|_{r_0,v}&=
\frac{\partial}{\partial r}S_m(\lambda_i,r)\Big|_{r_0,v}=0,\\
\label{eq:eomrhoonshell}
\rho\frac{\partial}{\partial\rho}\Gamma(\lambda_i, r, \rho/\mu)\Big|_{r_0,v}&=B(\lambda_i,r)\Big|_{r_0,v}=0.
\end{align}
\end{subequations}
These results are exact to all orders in the loop expansion. The form of these equations suggests a two-step approach to finding extrema: (1)~ Since $S_m(\lambda_i,r)$ 
is independent of $\rho$, the first equation 
\eqn{eq:eomronshell} demonstrates that the value $r_0(\mu)=\phi^2/R=\xi(\mu)/\lambda(\mu)$ of the ratio at an extremum can be inferred in tree approximation, although we do not know the scale $\mu$ at which the couplings are to be evaluated. (2)~The second equation 
\eqn{eq:eomrhoonshell} then determines the scale $\mu=v$ and 
the value of the curvature $\rho=v$, expressed as a special relationship 
among the couplings that must obtain at that scale.

In order to determine stability, we shall also need the matrix of second derivatives on-shell :
\begin{subequations}
\label{eq:secondonshell}
\begin{align}
\frac{\partial^2}{\partial r^2}\Gamma(\lambda_i, r, \rho/ \mu)\Big|_{r_0,v}&=
\frac{\partial^2}{\partial r^2}S_m(\lambda_i,r)\Big|_{r_0,v},\\
\rho\frac{\partial^2}{\partial r\partial\rho}\Gamma(\lambda_i, r, \rho/ \mu)\Big|_{r_0,v}&=
\frac{\partial}{\partial r}B(\lambda_i,r)\Big|_{r_0},\\
\rho^2\frac{\partial^2}{\partial \rho^2}\Gamma(\lambda_i, r, \rho/ \mu)\Big|_{r_0,v}&=
C(\lambda_i,r_0).
\end{align}
\end{subequations}
Given our conventions, these equations \eqn{eq:secondonshell} are also exact to all orders in the loop expansion, but their leading nonzero contributions vary from tree level for those involving $S_m$, 
to one-loop for $B$, to two-loop\footnote{As before, the two-loop contribution to $C_2$ can be calculated from one-loop corrections; $C_3$, 
from two-loop corrections, etc.} for $C$. 
The second variation on-shell is therefore
\beq\label{eq:deta2onshell}
\delta^{(2)}\Gamma=\frac{1}{2}
\begin{pmatrix}
\frac{\delta\rho}{\rho} & \delta r
\end{pmatrix}
\begin{bmatrix}
C(\lambda_i,\!r_0) &\ B{}^\prime(\lambda_i,\!r_0)\\
B{}^\prime(\lambda_i,\!r_0) &\ S_m^{''}(\lambda_i,\!r_0)
\end{bmatrix}
\begin{pmatrix}\frac{\delta\rho}{\rho} \\ \delta r
\end{pmatrix}.
\eeq
This matrix has two eigenvalues $\varpi_i$ that may be approximated as
\beq\label{eq:eigenvalues}
\varpi_1(r_0,v)=\frac{ S_m^{\,''}}{2}+O(\hbar^2),\qquad 
\varpi_2(r_0,v)=\half\left[C_2-\frac{\left(B'_1\right)^2}{S_m^{''}}\right]+O(\hbar^3).
\eeq
So $\varpi_1=\lambda(v)/2$ is determined by the classical curvature, 
and $\varpi_2$, although of order 
$\hbar^2$, by one-loop results, just as with $C_2$.

To flesh this out, we need to determine $B$ and $C$ from the RGE\footnote{As before, we shall suppress possible gauge parameters. The only gauge-dependent quantities here are the wave function renormalizations, which we shall show do not contribute to observables.}:
\beq\label{eq:rge3}
-\left[\mu\frac{ \partial}{\partial \mu}-\gamma_\rho\frac{\partial}{\partial\rho}\right]\Gamma(\lambda_i,r,\rho/\mu)
=\left[\beta_{\lambda_i}\frac{\partial}{\partial \lambda_i }-\gamma_r r\frac{\partial}{\partial r} \right]
\Gamma(\lambda_i,r,\rho/\mu).
\eeq
As in the preceding section, the left-hand side may also be expressed as $(1\!+\gamma_\rho)\rho\,\partial\Gamma/\partial\rho$. The first variations \eqn{eq:eomonshell} vanish on-shell, so, to all orders,
\beq
\beta_{\lambda_i}\frac{\partial}{\partial \lambda_i}\Gamma(\lambda_i,r,\rho/\mu)\Big|_{r_0,v}=0,
\eeq
for arbitrary $\mu$. To one-loop order, \eqn{eq:rge3} becomes
\beq
B_1+C_2\log(\rho/\mu)=\left[\beta_{\lambda_i}^{(1)}\frac{\partial}{\partial \lambda_i }
-\gamma_r^{(1)} r\frac{\partial}{\partial r}\right]\Big(S_{ho}(\lambda_i)+S_m(\lambda_i,r)+B_1\log(\rho/\mu)\Big),
\eeq
so that
\begin{subequations}\label{eq:oneloopoffshell}
\begin{align}
B_1(\lambda_i,r)&=\beta_{\lambda_i}^{(1)}\frac{\partial}{\partial \lambda_i }\left[S_{ho}(\lambda_i)+S_m(\lambda_i,r)\right]
-\gamma_r^{(1)} r S'_m(\lambda_i,r), \\
B'_1(\lambda_i,r)&=\beta_{\lambda_i}^{(1)}\frac{\partial}{\partial \lambda_i }
S'_m(\lambda_i,r)-\gamma_r^{(1)}\frac{\partial}{\partial r}\big(rS'_m(\lambda_i,r)\big), \\
C_2(\lambda_i,r)&=\left[\beta_{\lambda_i}^{(1)}\frac{\partial}{\partial \lambda_i }
-\gamma_r^{(1)} r\frac{\partial}{\partial r}\right]B_1(\lambda_i,r).
\end{align}
\end{subequations}
As claimed, $C_2$ is determined by one-loop results. Taking note of  \eqn{eq:eomonshell}, these become on-shell
\begin{subequations}\label{eq:onelooponshell}
\begin{align}
\label{eq:b1ronshell} 
B_1(\lambda_i,r_0)&=\beta_{\lambda_i}^{(1)}\frac{\partial}{\partial \lambda_i }\big[S_{ho}(\lambda_i)+S_m(\lambda_i,r)\big]\big|_{r_0,v},\\
\label{eq:b1rprimeonshell}
B'_1(\lambda_i,r_0)&=\beta_{\lambda_i}^{(1)}\frac{\partial}{\partial \lambda_i } S'_m(\lambda_i,r)\Big|_{r_0,v}
-\gamma_r^{(1)} r_0 S^{\,''}_m(\lambda_i,r_0), 
\\
\label{eq:c2ronshell}
C_2(\lambda_i,r_0)&=\beta_{\lambda_i}^{(1)}\frac{\partial}{\partial \lambda_i }B_1(\lambda_i,r)\Big|_{r_0,v}
-\gamma_r^{(1)} r_0B'_1(\lambda_i,r_0).
\end{align}
\end{subequations}
Thus, on-shell, $B_1$ is independent of $\gamma_r$, but $C_2$ is not. This reflects the fact that, 
from \eqn{eq:eomrhoonshell}, the condition $B_1=0$ is one of the (leading order) 
conditions for an extremum. From \eqn{eq:eigenvalues}, on the other hand, we see that (unlike 
in the pure gravity case discussed in the last section) the sign of $C_2$ does not determine the 
nature of the extremum; we must calculate $\varpi_2$. We find 
\beq\label{eq:varpi2}
\varpi_2=\half\left[\left(\beta_{\lambda_i}^{(1)}\frac{\partial}{\partial \lambda_i }\right)^{\!2}\! \big[S_{ho}(\lambda_i)+S_m(\lambda_i,r)\big]
-\frac{1}{S_m^{''}}\left(\!\beta_{\lambda_i}^{(1)}\frac{\partial}{\partial \lambda_i } S'_m(\lambda_i,r)
\!\right)^{\!2}  \right]\Big|_{r_0,v}.
\eeq
The order of operations is important; the derivatives with respect to
the couplings must be carried out before setting $r\!=\!r_0$. Note that
the $\gamma_r$-dependence has cancelled out between the two terms in
\eqn{eq:eigenvalues}, as we anticipated, because the result \eqn{eq:varpi2}  
must be gauge invariant.

There is still quite a lot of work to be done to evaluate and solve
\eqns{eq:eomronshell}{eq:eomrhoonshell} for potential extrema and to
evaluate \eqn{eq:varpi2} to determine local stability. First of all, we
need the one-loop beta-functions. These have been given several places
in the literature and, for easy reference, are reviewed in
Appendix~\ref{sec:betas} in the present notation. Quite generally, we
see that  $\beta_a$ and $\beta_\varepsilon$ have the same form as in
\eqn{eq:betahoae}, but with the positive constants $\beta_2$ and
$\beta_1$ dependent upon the matter content. In this model with one real
scalar only, $\beta_2=799/60$ and $\beta_1=523/120$.  Thus, the coupling $a$
is always AF and, at one-loop, $\beta_a$ is independent of the other
coupling constants, a residue of the conformal invariance of the Weyl
tensor. Noting that $a_0/a=1+a_0\beta_2 t$, where $dt = \kappa d(\ln \mu)$, 
it proves useful to define a
new parameter
\beq 
u\equiv (1/\beta_2)\log(a_0/a)=(1/\beta_2)\log(1+a_0\beta_2 t),
\eeq
so that $du=a dt = -da/(\beta_2 a).$ 
The coupled equations simplify considerable if we rescale 
$\lambda$ as we did with $b$, $y\equiv \lambda/a$. Then the three remaining variables $w,\xi,y$ obey
\begin{subequations}
\label{eq:betabarmat}
\begin{align}
\frac{dw}{du}\equiv\overline{\beta}_w&=\frac{10}{3}\left[ 
w^2-\frac{1099}{200}w+\frac{1}{8}
+\frac{1}{5}\left( \frac{6\xi+1}{4} \right)^2 \right]; \\
\frac{d\xi}{du}\equiv\overline{\beta}_\xi&=\left[\left(6\xi\!+\!1\right)y\!-\!\xi
\left( \frac{3\xi^2}{2}\!+\!4\xi
\!-\!3\!+\!\frac{10\,w}{3}\!-\!\frac{1}{4w}\left(9\xi^2 \!+\!20\xi\!-\!4\right) 
\!\right)\right];\\
\label{eq:betabarmatC}
\frac{dy}{du}\equiv\overline{\beta}_y&=\left[18y^2\!+\!y\left(\!\frac{499}{60}\!-\!3\xi^2
\!+\!\frac{1}{2w}\left(1\!+\!12\xi\!+\!33\xi^2 \right)\!\right)\!+\!
\frac{\xi^2}{2}\!\left(\!5\!+\! \frac{(6\xi\!+\!1)^2}{4w^2} \right)\!\right]\!.
\end{align}
\end{subequations}
Note that $a$ no longer appears in these ``reduced" beta-functions. This suggests that there may well be fixed points at finite $w,\xi,y$, where all three beta-functions simultaneously vanish. 

Although our primary interest is in finding where
$B_1(\lambda_i,r_0)=0,$ let us first explore whether there are fixed
points. First, note that, if $\xi=0$ (minimal coupling), then
$\overline{\beta}_\xi=0$ implies $y=0$ as well. $\overline{\beta}_y$
also vanishes for $\xi=y=0.$ Then $\overline{\beta}_w=0$ implies
$w\approx0.02514$ or $w\approx5.46986.$ Other fixed points are more
difficult to locate and must be found numerically, but we found four
more. All the fixed points are shown in Table~1.

\begin{table}[ht]
\begin{center}
\begin{tabular}{|c|c|c| c| c| } \hline
& $ w $ & $ \xi $ & $ y $ & $ z $\\ \hline
$1.\ $ & $ 0.02514 $ & $ 0. $ & $ 0. $ & $ n.a. $ \\ \hline
$2.\ $ & $ 0.36011 $ & $ 1.7907 $ & $ -4.8714 $ & $ -1.3710 $ \\  \hline
$\!{\bf 3.} $ & ${\bf 0.02450} $ & $ {\bf-0.02519} $ 
& $ {\bf -1.2726} $ & ${\bf -0.015265 } $ \\ \hline
$4.\ $ & $ 0.03336 $ & $ 0.1898 $ & $ -0.2643 $ 
& $ -3.0652 $ \\ \hline
$5.\ $ & $ 5.4699 $ & $ 0. $ & $ 0. $ & $n.a. $ \\ \hline
$6.\ $ & $ 5.4705 $ & $ -0.02567 $ & $ -0.4654 $ & 
$ -1.941\!\times\!10^{-4} $ \\  \hline
\end{tabular}
\caption{\label{FP}Fixed Points}
\end{center}
\end{table}
\noindent Of the six, it can be shown that all are saddle points except for
the one located at $w\!\approx 0.0245 , \xi\!\approx\!- 0.0252 ,
y\!\approx\! - 1.273$, which is UV attractive. Unfortunately, since
$y<0$ near there, it has the opposite sign to the one required for stability
in $r$, \eqn{eq:offshellreal}.  So this does not appear to be an 
acceptable model for large scales.  Further, as we shall 
explain in Section~\ref{sec:constraints}, no renormalization trajectory 
can cross from $y>0$ to $y<0$.  (This would require a change in sign of 
the curvature, so it is not surprising that it is different phase.)

\begin{figure}[ht]
\epsfysize=10cm \centerline{\epsfbox{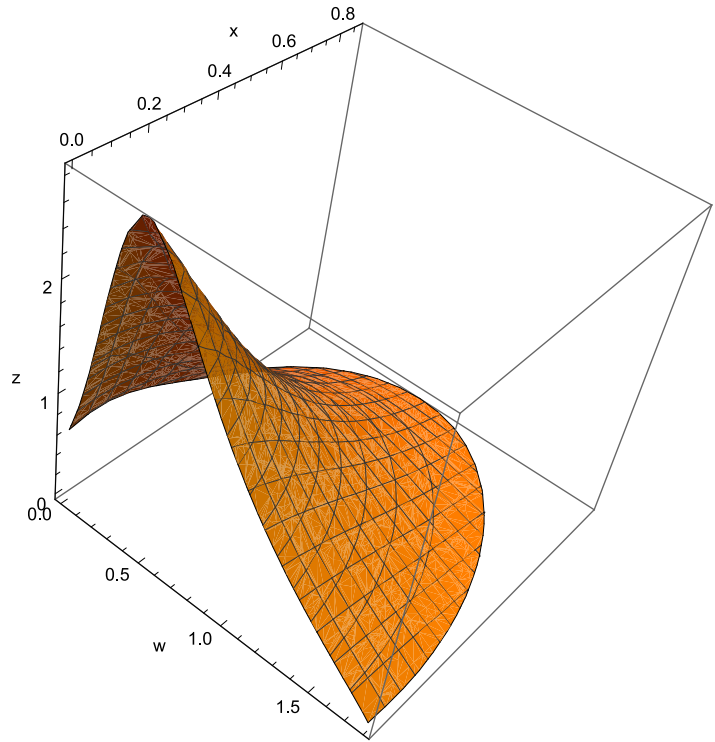}} \caption{$B_1=0$ for $w>0$.} \label{fig:A}
\end{figure}

Returning to the determination of $B_1,$
from \eqn{eq:offshellreal}, we have the values of the 
matter action and its derivatives and, as noted previously, 
$r_0=\xi/\lambda$, which implies $a(\mu)r=\xi(u)/y(u)$. 
From \eqn{eq:b1ronshell}, we may write the on-shell value 
\begin{align}
\label{eq:b1realscalar}  
B_1(\lambda_i,r_0) &=\beta_a\frac{\partial}{\partial a}[S_{ho}+S_m]+\beta_w\frac{\partial}{\partial w}S_{ho}
+\left[\beta_\xi\frac{\partial}{\partial \xi}+\beta_y\frac{\partial}{\partial y}\right]S_m,\\
\hbox{or}\quad \frac{B_1}{\kappa V_4}&=\frac{\beta_2}{6}\left[2w-\frac{\beta_1}{\beta_2}-\frac{3\xi^2}{2y}\right]+\frac{1}{3}\overline{\beta}_w
+\frac{\xi}{4y}\left(\frac{\xi}{y}\overline{\beta}_y - 2\overline{\beta}_\xi\right),
\end{align}
where the $\overline{\beta}_i$ are given in \eqn{eq:betabarmat}. From their form, one can observe that $B_1=B_1(w,\xi,y)$ has no explicit dependence on $a$. The equation to be solved, $B_1=0$, is of the form $P(w,\xi,y)/w^2y^2$, 
where $P$ is a polynomial in the three variables with highest degree $w^4\xi^4y^2$. Not surprisingly, there is a continuum of (real) solutions satisfying $P(w,\xi, y)=0$. Because the classical action on-shell depends on $\xi,y$ only through the ratio $\xi^2/y$, this space of solutions is more easily represented in terms of different variables. Changing from $y$ to $z$ with $z\equiv 3\xi^2/(4wy)$, we find that 
\begin{align}\label{eq:b1z}
\frac{B_1}{\kappa V_4}&\!=\! \left(\frac{20 w^2 + (1 + 6 \xi)^2}{18}\right)\! (z+1)^2\!+\!
\frac{z}{3}\left(\! 8 \xi (w\!-\!1) \!-\!11w \!+ \frac{13}{6} \right)\!-\xi\!-\!\frac{5w}{3}\!-\!\frac{151}{240},\!
\end{align}
which is only quadratic in each parameter and non-degenerate in $z$,  
since, as discussed earlier,  $w\ne0$ in perturbation theory. The
contour plot of $B_1=0$ is a very large region, much of which is not of
particular physical interest.  As remarked below \eqn{eq:offshellreal},
we may restrict our search to $\lambda>0$, \ie, $y>0$.  Therefore, the
signs of $z$ and $w$ must agree. Further, in order to recover Einstein
gravity below the DT scale, we must have $\xi>0$. Although we allow $w$
to have either sign, it is most convenient to display the contour plot
of perturbative solutions for the regions $w>0$ and $w<0$ separately. 
For $w>0,$ we find the contour plot of solutions in Fig.~\ref{fig:A}. 
It is noteworthy that the entire space of solutions lies within the
limits $0<w\lesssim1.79$, $0<\xi\lesssim0.862$, and $0<z\lesssim2.82.$  

\begin{figure}[ht]
\epsfysize=10cm \centerline{\epsfbox{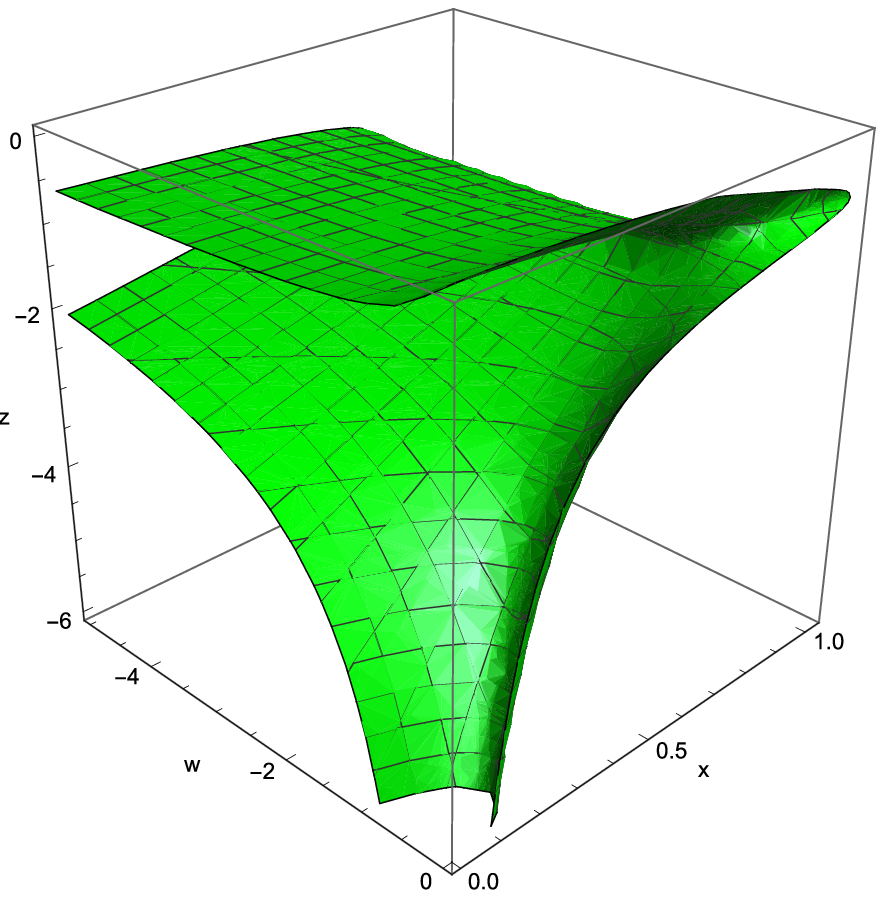}} \caption{$B_1=0$ for $w<0$.} \label{fig:C}
\end{figure}

\begin{figure}[ht]
\epsfysize=10cm \centerline{\epsfbox{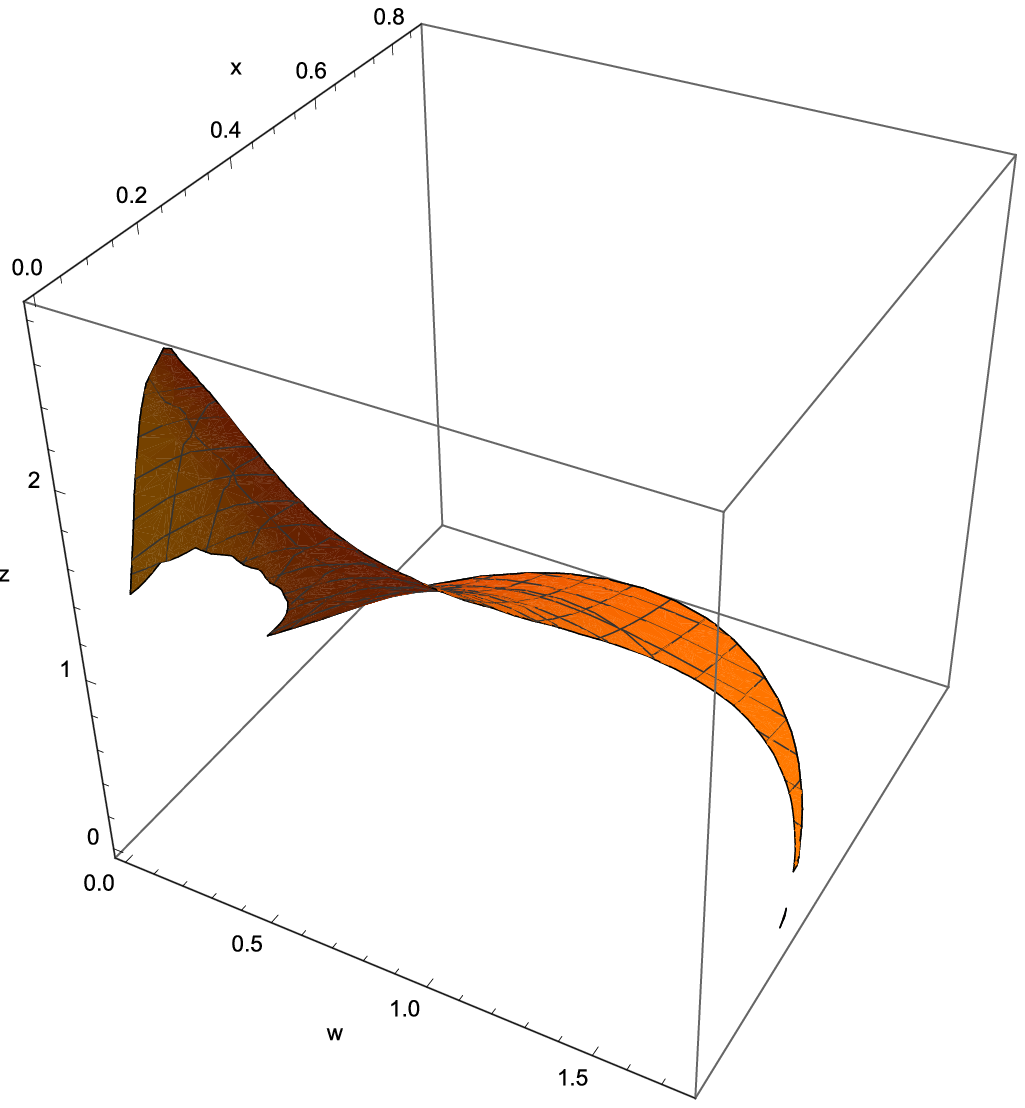}} \caption{$B_1=0$ for $w>0$, $\varpi_2 > 0$.} \label{fig:B}
\end{figure}

\begin{figure}[ht]
\epsfysize=10cm \centerline{\epsfbox{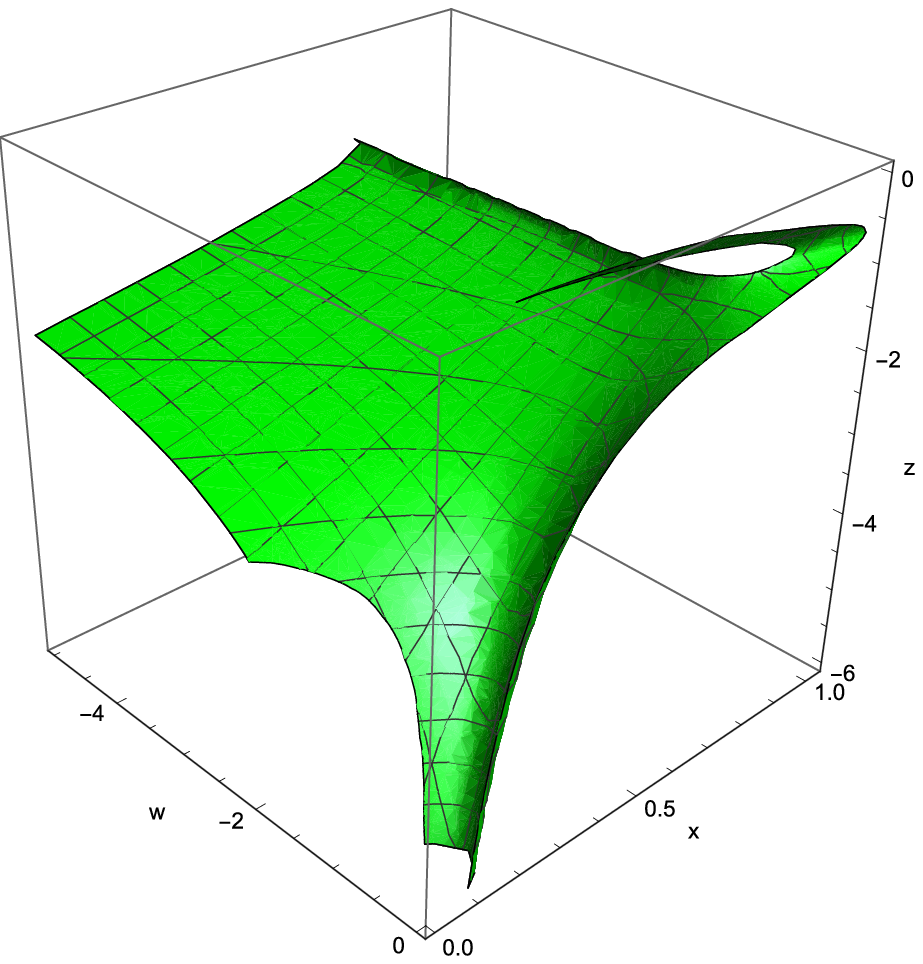}} \caption{$B_1=0$ for $w<0$, $\varpi_2 > 0$.} \label{fig:D}
\end{figure}

For $w<0,$ the contour plot is more complicated, especially because the
equation $B_1=0$ degenerates at $z=-1,$ where it becomes simply $6 w + 5
\xi - 8 w \xi=973/240.$   For all $w<0,$ this has solutions
$\xi=(1/240)(973 - 1440 w)/(5-8w)$, so that $\xi$ lies within the fairly
narrow range $3/4 < \xi<0.811.$  A portion of the general  contour plot
for $w,z<0,\xi>0$ is shown in Fig.~\ref{fig:C}.

The next step is to determine the subspace of the preceding solutions to $B_1=0$ that are local minima, \viz, those having $\varpi_2>0,$ in \eqn{eq:varpi2}.  The explicit expression is straightforward to calculate but messy:
\begin{align}
\varpi_2(w,\xi,z)&=\frac{\kappa^2 a}{4320wz}\left[
    540 \xi ^2 \left(-1+6 \xi +72 \xi ^2\right) 
  \right.\cr
&\left. \hskip-17mm 
 +10 z \Big(1600w^4\!-9992w^3\!+2 w^2\! \left(3407 + 600 \xi\! - 2160 \xi^2\right) \right.  \cr
&\left.\hskip-14mm 
+18 \xi \! \left(19-149 \xi\!  +375 \xi ^2\!+324 \xi ^3\right)\!
 -3 w\! 
 \left(55 + 528 \xi\! - 2400 \xi^2\! + 1116 \xi^3\! + 432 \xi^4\right)\!\!\Big)\right. \cr
&\left. \hskip-17mm 
-z^2 \Big(\!16000 w^4+160 w^3 (289+600 \xi )
+4 w^2 \left(2363 - 33624 \xi + 71280 \xi^2 + 2160 \xi^3\right)
\right.\cr
&\left. 
\hskip-7mm +1080 w \left(-45 + 112 \xi - 344 \xi^2 + 176 \xi^3 + 24 \xi^4\right)\right.\\
&\left.\hskip-3mm 
 +15 \left(355-1204 \xi +5820 \xi ^2-10800 \xi ^3+5184 \xi ^4\right)\!\Big)\right.\cr
&\left. \hskip-17mm 
-10 z^3 \Big(\!101+36 \xi(34 + 74 \xi - 105 \xi^2 + 414 \xi^3) 
+ 4\left(1200 w^4 + 2w^3 (-191 + 720 \xi)\right.\right.\cr
&\left.\left.\hskip-3mm
+10 w^2 (53 + 36 \xi + 576 \xi^2) + 
 3 w(1+6\xi) (-43 - 270 \xi + 192 \xi^2+ 18 \xi^3)\right)\!\Big)
\right.\cr
&\left. \hskip-15mm
-40 z^4\! \left(20 w^2+(1+6 \xi )^2\right)^2    \right]\!.\nonumber
\end{align}
Since we require  $y>0$, it follows that $wz>0$,  so the
polynomial\footnote{It is of fourth-degree in each of the three
parameters $w,\xi,z$ except at $z=-1$, where it becomes cubic in
$w,\xi.$}  in brackets must be positive for the extremum to be a local
minimum.  The intersection of the region $\varpi_2(w,\xi,z)>0$ with the
$B_1(w,\xi,z)=0$ surface is shown in Fig.~\ref{fig:B} for $w>0$ and in
Fig.~\ref{fig:D} for $w<0.$

Therefore, we have shown that there remains a continuum of local minima at which DT takes place.  All such points are candidates for no-particle solutions (vacua) in this model.
To illustrate, some values for $w>0$, for $z\approx .0005,$ one has $B_1=0$ 
and $\varpi_2>0$ for $0 < w < 1.78$ with  $\xi =0.0834 - 0.000333 w + \sqrt{0.293 + 0.833 w -
0.556 w^2}>0$. At the other extreme, for $z\approx 2.82,$ one finds 
$0.357 < w < 0.364,$ with $\xi =-0.0207 - 0.129 w + \sqrt{-0.0756 +
0.417w - 0.539 w^2}>0.$  

As a renormalizable completion of Einstein gravity, this model is unsatisfactory for a reason that is not immediately apparent.  The only UV fixed point has $y<0,$ whereas all local minima must have $y>0.$  One can show that as the couplings evolve from lower to higher scales, no path runs from $y>0$ to $y<0.$  Therefore, the region of parameter space in which DT occurs is not connected to the region in which AF holds.  Invariably, one or another of the couplings grows and perturbation theory breaks down. (We have not investigated whether calculable nonperturbative effects, such as instantons, might alter this conclusion, but it seems doubtful.)  This property does not appear to be a generic property of any such model, and we can hope (along with previous authors~\cite{Fradkin:1981iu}) that a richer theory of matter, 
such as a grand unified theory, might avoid such a conclusion. 

\section{Additional Matter} \label{sec:morematter}

\paragraph{ }
It is straightforward to generalize the model above to a more
general theory  containing gauge, Yukawa and additional scalar
multiplets if we assume that the scalar sector we have described is a
hidden sector, interacting with what we may call the matter sector  only
via gravitational interactions. This is because at one-loop order, the
$\beta$-function and effective  potential calculations we have described
are unaffected, except for matter contributions  to the
$\beta_{a,b,\varepsilon}.$ (We assume here that the dominant 
non-minimal $\xi$-type coupling at the DT scale 
is to the original $\phi$-singlet.)
We reproduce these generalized 
$\beta$-functions in appendix~\ref{sec:betas}. 
Generally speaking the results remain
qualitatively the same.  For example, with $N_0 = 5, N_{{1}/{2}} =
24, N_1 = 0, N_{1}^{0} = 12$, corresponding to a coupling of our theory to the 
Standard Model (including right-handed neutrinos),  we find 

\begin{table}[ht]
\begin{center}
\begin{tabular}{|c|c|c| c| c| } \hline
& $ w $ & $ \xi $ & $ y $ & $ z $\\ \hline
$1.\ $ & $ 0.0271 $ & $ 0. $ & $ 0. $ & $ n.a. $ \\ \hline
$2.\ $ & $ 0.7497 $ & $ 1.2518 $ & $ -0.7150$ & $ -2.1926 $ \\  \hline
$\!{\bf 3.} $ & ${\bf 0.0242} $ & $ {\bf-0.0285} $ 
& $ {\bf -1.5129} $ & ${\bf -0.0166 } $ \\ \hline
$4.\ $ & $ 0.0567 $ & $ 0.1771$ & $ -0.1103 $ 
& $ -3.764 $ \\ \hline
$5.\ $ & $ 6.9280 $ & $ 0. $ & $ 0. $ & $n.a. $ \\ \hline
$6.\ $ & $ 6.9309 $ & $ -0.0299 $ & $ -0.7348 $ & 
$ -1.318\!\times\!10^{-4} $ \\  \hline
\end{tabular}
\caption{\label{FPSM}Fixed Points with Standard Model matter}
\end{center}
\end{table}

The fixed point with (now) $w \approx 0.0242$ remains UV attractive,
although  two of the eigenvalues of its stability matrix develop
imaginary parts. This simply means that the couplings oscillate around
an envelope that is AF.

As in our original model, there is a substantial range of parameter 
space such that $B_1 = 0$ represents a perturbatively stable minimum  of
the effective potential. Thus it is  feasible to entertain the
possibility that a realistic theory might  be constructed with a ``hidden
sector" responsible for generating the  Planck mass via DT. 

Of course if we wished to take seriously the above possibility in the 
SM context, we would need to consider the indication of new physics 
associated with the electroweak vacuum stability issue, caused  by the
running to negative values of the Higgs quartic coupling $\lambda_H
(\mu)$. A recent comprehensive
analysis~\cite{Buttazzo:2013uya} suggests the possibility that a new
physics threshold is required  at a scale of around $\Lambda_I \approx
10^{10} - 10^{12}$~GeV~\footnote{Note that while $\lambda_H (\mu)$ is of
course gauge invariant for all  $\mu$, defining the instability scale
by, for example, $V(\Lambda_I) = 0$  is manifestly gauge dependent, so
care is required\cite{Andreassen:2014gha},\cite{DiLuzio:2014bua}.}. 
Now the DT scale in our model is given by 
\beq
\Lambda_{\rm DT} \sim \sqrt{R} \sim \sqrt{\frac{\lambda}{\xi}}\vevof\phi \sim 
{\sqrt{\lambda}M_P}/{\xi},
\eeq
 so to make this scale coincide with
$\Lambda_I \sim 10^{12}$~GeV requires ${\sqrt{\lambda}}/{\xi}
\sim 10^{-7} - 10^{-9}$. 
Proponents of Higgs inflation~\cite{Bezrukov:2007ep, Barvinsky:2008ia, 
DeSimone:2008ei, Bezrukov:2009db, Barvinsky:2009fy} are
content to contemplate large values of $\xi$ ($\xi \approx 10^4$), but it
is clear from, for example \eqn{eq:betabarmat}, that such $\xi$ values
lead to  loss of perturbative credibility for our calculations.

Evidently it will also be interesting to entertain a more complicated 
generalization where the non-minimally coupled scalar sector 
has gauge and Yukawa interactions. For example, one could imagine 
a Grand Unified Theory (GUT) 
wherein the DT-generated vev for the scalar fields both generated Einstein gravity 
and broke the  GUT gauge invariance down to the SM.  We postpone this possibility 
for future discussion. 

\section{Constraints on Coupling Constants}\label{sec:constraints}

\paragraph{ }
What constraints exist on the couplings constants? First of all, unlike
flat space field theories, not all spacetimes can be analytically
continued from Lorentzian to Euclidean signature.  We tacitly  assume
that all physically realizable spacetimes arise by the reverse process
of continuation from a Euclidean metric. We are especially interested in
models in which the couplings are asymptotically free so that
perturbation theory can be used to determine the solutions. We have
already discussed some properties of the effective action at the DT
scale in two cases, the pure $R^2$-model of gravity and the $R^2$ plus a
real scalar.   We also touched on inclusion of the SM fields in a hidden
sector.  

In the case of no matter, we found that there were no local minima of
the effective action, regardless of the signs of the couplings.

In the case of the real field and its simple extension discussed in
Section~\ref{sec:morematter}, the basin of attraction of the AF fixed
point is a distinct phase from the range of parameters where DT occurs. 
This can be seen as follows:  We required $y>0$ at the DT scale for
stability.  If the couplings are to approach the AF fixed point where
$y<0,$ then the trajectory as some point will have to cross $y=0.$ If at
some point $y\to0$ for positive $y$, then we see from 
\eqn{eq:betabarmatC} that $\overline{\beta}_y>0$ for all values of
$\xi,w$, so $y$ must increase from such a point. Therefore, $y$ cannot
become negative, at least, not so long as perturbation theory is valid.

There has been considerable discussion in the
literature\footnote{See Chapter 9 of \reference{Buchbinder:1992rb}\ for a
summary of some models.} of whether AF for all couplings obtains, but we
have not seen previous discussions of whether or not the couplings
actually run from their on-shell values to their AF values.  Our result
appears to be model-dependent, and there can be hope that
this obstruction will be remedied in future, more realistic models. 
Nevertheless, this is an issue that requires attention, even in
the existing models.

We found in Appendix~\ref{sec:stability} that the EPI for $\Delta\Gamma$
was convergent only if all couplings $a,w,\xi,y$ are positive with
$w<3/2+3\xi^2/(4y)$.  We postponed the question of whether this is
required to be true at all scales or, in particular, at the DT scale
$\mu=v.$  We believe the answer in both cases is ``no", based on
experience with flat space models.  First, consider the familiar double-well potential
$V=\lambda\phi^4/4-m^2\phi^2/2$ with $\lambda,m^2>0,$ we know that it is
stable near the classical minima $\pm v$.  Between the two minima, the
true effective potential is simply a straight line between the two
minima, but the perturbative effective potential  resembles the
classical potential.  The one-loop correction to the effective potential
is 
\beq
\Delta V^{(1)}=\frac{(3\lambda\phi^2-m^2)^2}{64\pi^2}\log(3\lambda\phi^2-m^2).
\eeq
This does have physical meaning in certain situations~\cite{Weinberg:1987vp}, 
even when $\Delta V^{(1)}$ becomes imaginary.  The imaginary part 
represents half the decay rate per unit volume, as expected, 
although the decay process is rather complicated.  Nevertheless, unstable modes do not necessarily invalidate the perturbative result when properly interpreted.

Second, consider the case of DT in massless scalar electrodynamics~\cite{Coleman:1973jx}, 
which is a model that is not AF.  In general, we
believe that the self-coupling of the scalar field $\lambda(\mu)$ must
be positive for the convergence of the EPI and for the potential to be
bounded below. However, if one adopts a renormalization scheme similar
to the one used here, the self-coupling $\lambda(\mu)$ turns negative at
the DT scale~\cite{Yamagishi:1981qq}. This is permissible because
$\lambda(v)$ is unusually small at the minimum $v$, comparable in size
with the electromagnetic one-loop correction; $\lambda(v) \sim - \kappa
\alpha^2(v)$. At somewhat lower scales, $\lambda(\mu)$ is
positive, typically on order of $\alpha,$ and larger than the one-loop correction.  
However, at very small or very large scales,
it becomes large, and perturbation theory breaks down.

With these cautionary examples in mind, let us consider the immediate
applications in this paper. There are many well-known
problems~\cite{Carlip:2001wq} defining functional integrals, especially
when gravity is included.  Among them is that the manifold over which
one integrates and the determination of the metric are intertwined, so
we seem to be caught in a vicious circle.   Further, in our case, the
action~\eqn{eq:lho2} includes the ``topological term" $G$ and possibly
also boundary or surface integrals. Beyond perturbation theory, we have
little to add to these issues.  Within perturbation theory, it seems
that the effects of the topological term $G$ can be restricted to the
``classical" background, and we do not need to address the topology of
the quantum fields.   In the background field method, summarized in
Appendix~\ref{sec:bfm}, it is required to evaluate the auxiliary
functional~\eqn{eq:kgen}.  This will converge if  the source-free EPI
$\Delta \gcal[\phi_i,0]$  converges, since $\Delta S$ is at least
quadratic in the  quantum fields\footnote{Fermion fields can be included
 without changing the basic results, but they would require a separate
discussion.}.

For the matter-free case in Section~\ref{sec:hogravity}, it is necessary
to find a scale where $a>0$ and $0<w(\mu)<3/2$ in order to evaluate the
integral, and that is possible.  Now if $a>0$ at one scale, its sign
cannot change (so long as perturbation theory holds.)  For $0<w<3/2$,
the EPI must agree with our determination of $B_1$ and $C_2$ in
\eqns{eq2:b1ho}{eq2:c2ho} via the RGE.  For $w(\mu)$ outside this range
of values, the effective action $\Gamma$ will develop an imaginary part,
as evidenced by unstable modes in the EPI.  Solving the equation
$B_1=0,$ we found extrema at  $w_-\approx -0.3$ and $w_+\approx 1.8$,
both outside the range of convergence of the EPI.  Accordingly, our RGE
calculation gives only the real part of $\Delta\Gamma$ and does not tell
us that there is an imaginary part as well.  However, we did determine
that  both extrema were local maxima, $C_2<0$, so we should expect
$\Delta\Gamma$ to have an imaginary part.  It is not given simply by
replacing $\rho$ by $-\rho$ in the logarithms in  \eqn{eq:gloops}.  In
terms of the eigenmodes of the Laplacian outlined in
Appendix~\ref{sec:stability}, the coefficient of the imaginary part will
come from only those modes that are negative.  Since all modes for
sufficiently large $n$ are positive, there are only a finite number of
unstable modes, so that this is consistent with renormalizability.

For the real field in Section~\ref{sec:realscalar}, we required $y>0$
for stability of the ratio $r=\phi^2/R.$  The constraints on the tensor
sector for convergence of the EPI are  $a>0$, $w<3/2+3\xi^2/(4y).$  (See
\eqn{eq:tensormodes}.)   From the conformal sector, \eqn{eq:conformal},
we must have $w>0$ and $\xi>0.$ We also learn that there are six zero
modes,  one, the dilaton, associated with SSB of the classical scale
invariance, and the other five as in the matter-free case, associated
with $\varphi_1$.  As explained in Appendix~\ref{sec:stability}, the
dilaton will get a mass$^2$ at two-loop order proportional to $C_2$.  We
do not know what will happen to the other five modes, whose origin
remains obscure to us.  

Although we found an AF fixed point, it has $y<0$ and $\xi<0$, outside the bounds above.  Thus, independently of the existence of DT, the EPI does not converge for values of the couplings near the AF fixed point!  That is decidedly unsatisfactory for a perturbative solution to exist.  

Are these inequalities also necessary at the DT scale?  We did find
once again that $a>0$ and $y>0$.  We also required $\xi>0,$ so that the
gravitational constant has the correct sign.  On the other hand, it is 
not clear that we must have $w(v)>0.$   Within these restrictions, we
found a large region of parameter space where DT can occur ($B_1=0$) and
where the points are local minima.  Thus, this is a viable mechanism for
generation of the Planck scale.  Which of these candidate vacua might be
acceptable would require a cosmological analysis, but this is not a
realistic model anyway.
  
In summary, the question of constraints on the couplings is thus scale
dependent and depends on the phenomena of interest.  However, to
calculate radiative corrections, one may make the separation between the
classical background and quantum corrections at any convenient scale
and later determine whether the questions of interest are at scales at
which the couplings are still small.  Especially for AF theories,
starting at a very large scale where the couplings are small is an
attractive possibility, but, unlike the models in this paper, we would
want the EPI over the quantum field converge in that domain.

Unfortunately, in the examples studied in this paper, one or another
physical or aesthetic requirement was violated. For the pure metric
model, Section~\ref{sec:hogravity}, the extrema were maxima rather than
minima.  With the addition of a real scalar, we found that we were able
to find a purely real radiative correction $\Delta\Gamma$ for values of
the couplings where there were local minima of the effective action, but
this region turned out not to be continuously connected to the AF
domain, so that, above the Planck scale, these theories appear to become
strongly coupled.  Whether they must be regarded as incomplete, we
cannot say, but this result is disappointing.  We do not know whether
the problem lies with our insistence on maximal isometry for the
background, but it would be rather surprising if the most symmetric
situation has problems not shared by less symmetric backgrounds.

This does not invalidate our conclusion that DT can occur perturbatively.  
At the DT-scale $v$, one-loop corrections are crucial, and
 the couplings need not respect the inequalities required near the AF
 fixed point. The only inequalities we can impose at the DT scale are those 
 required for stability at that scale, such as $y(v)>0$, $\varpi_2(v)>0$.

\section{Conclusions}\label{sec:conclusions}

\paragraph{ }
In this paper, we have presented a number of new formal results for
classically scale-invariant, renormalizable gravity models.  Such models
can be motivated by the fact that the scale breaking (due to the anomaly
of the corresponding QFT) is soft, preserving naturalness, unlike models
that include explicit scale-breaking in the action.   They are also
attractive, in that $R^2$-gravity is not only renormalizable but also
asymptotically free (AF), a quality generally preserved when
renormalizable interactions involving matter fields are added.

We extended the formalism for determining whether dimensional
transmutation (DT) takes place to include the background metric itself,
at least for maximally symmetric backgrounds.  We analyzed the situation
in the absence of matter, showing that there was not a
perturbative background that was locally stable.  

Classically, a scale-invariant theory that is spontaneously broken
yields a massless Goldstone boson.  Since DT is a form of spontaneous
symmetry breaking, there remains such a massless particle in lowest
order in perturbation theory in the QFT.  However, since  the QFT breaks
scale invariance due to the anomaly, this particle becomes  massive from
radiative corrections (that  are second-order in the loop-expansion.) 
We showed how this mass-squared could be determined from a certain
collection of the one-loop results, without having to face the daunting
task of computing the full two-loop corrections to the effective action.
 This allows one to calculate the local curvature at the DT scale, which
determines whether or not an extremum corresponds to a local minimum or
maximum of the effective action In a kind of corollary to the
discussion here, we elaborate in a companion 
paper~\cite{Einhorn:2014bka} how the effective action of this theory 
may be regarded in a sense as involving only two gravitational
couplings rather than three, and how this observation relates  to a
possible $a$-theorem for $R^2$ gravity.

The preceding observations and formulas can be extended to $R^2$-models
that include matter, as was illustrated by considering the simplest case
of the addition of a real scalar field. Despite its simplicity, several
properties not previously explored concerning $R^2$-models emerged. The
non-minimal coupling $\xi,$ and the ratios of couplings, $w=a/b$ and
$y=\lambda/a$, have a number of finite fixed points, only one of which
is UV attractive.   The basin of attraction of this AF fixed point is
limited, and,  in fact, does not include the region in which DT minima
occur.   Accordingly, the couplings in the regions where DT occurs are
not AF and become either strongly coupled or somehow modified at high
scales.  We did not attempt to determine the behavior at strong
coupling; it may depend on which couplings become large at high scales. 
Treating this as an effective field theory, we then showed that DT can
occur over a very wide range of parameters and that a large subset of
these extrema are in fact minima of the effective action.

In future work, we shall examine theories including other matter
fields.  Some models that included  non-Abelian gauge fields were
partially investigated in refs.~\cite{
Buchbinder:1992rb,
Elizalde:1994gv,
Elizalde:1995at}.  Some of these are
claimed to be AF in all their essential couplings, which is to be
welcomed.  If there are classically scale-invariant models of this type
 that undergo perturbative DT, they will have effective field theories
below the DT scale that look like Einstein gravity, at least so long as
they have a positive gravitational constant (non-minimal couplings $\xi$
greater than zero with our conventions.)  If such ``vacua" were within
the basin of attraction of a UV fixed point for an AF theory, then this
would be a candidate model for a unified theory of all interactions,
including gravity, in which the mass scales would be determined solely
by DT.  These are obviously very attractive candidates for further
exploration.  Whether any such model is consistent with unitarity
remains an unresolved issue, which can be addressed after
finding a model that is acceptable in other respects.

\section{Acknowledgements}
\paragraph{ }
One of us (MBE) would like to thank A.~Vainshtein and E.~Rabinovici for
discussions and to acknowledge the use of xAct~\cite{Martin-Garcia:2014www} and 
xTras~\cite{Nutma:2013zea}, and to thank J.~M.~Mart\'in-Garc\'ia and his team, as well 
as R.~McNees, for discussions and advice.    
DRTJ thanks I.~Jack for conversations, I.~Avramidi for correspondence, 
KITP (Santa Barbara) for hospitality and
financial support, and the Aspen Center for Physics, where part of this
work was done, for hospitality. This research was supported in part by
the National Science Foundation under Grant No. NSF PHY11-25915 and
Grant No. PHYS-1066293.
\medskip

\noindent {\it Note Added}

\indent After this paper was completed, our attention was drawn to
\reference{Salvio:2014soa}\  wherein DT in $R^2$ gravity with matter is
also proposed in the context of the Standard Model.   Their treatment
differs considerably from ours. We do not believe their results are
applicable  to de Sitter space, in particular, with regard to their
neglect of the Gauss-Bonnett term  and its associated $\beta$-function.
As they point out, some of their other $\beta$-functions  disagree with
those in the literature, which we used herein.  We thank A.~Salvio and
A.~Strumia for correspondence.

\begin{appendix}

\section{Gauss-Bonnet Relation} \label{sec:g-b}

\paragraph{ }
The local Gauss-Bonnet relation is that a linear combination of three
quadratic invariants, $C_{\kl\mn}^2, R_\mn^2$ and $R^2$ is, in
four-dimensions, a total derivative. It can be written in a variety of
ways~\cite{Fradkin:1985am}:
\begin{subequations}
\label{eq:gaussbonnet} 
\begin{align}
\label{eq:gb1}
& R^*R^*=R_{\kl\mn}^2-4R_\mn^2+R^2
=C_{\kl\mn}^2-2\widehat{R}_\mn^{\,2}+\frac{1}{6}R^2\!\equiv\! G,\\
\label{eq:gb2}
&\widehat{R}_\mn\!\equiv\!{R}_\mn-\frac{g_\mn}{4}R,\ \ 
R^*{}^{\kl\mn}\!\equiv\! \half\epsilon^{\kl\ab}R^\mn{}_\ab,\ \ 
R^*R^*=\frac{1}{4}\epsilon^{\kl\ab}\epsilon_{\mn\gamma\delta}R^\mn{}_\ab R^{\gamma\delta}{}_{\kl},\\\label{eq:gb3}
& R^*R^*=\nabla_\mu B^\mu,\ \ 
B^\mu\!\equiv\! \epsilon^{\mn\gamma\delta}\epsilon_{\rho\sigma}{}^\kl\Gamma^\rho_{\kappa\nu}\left[ \half R^\sigma{}_{\lambda\gamma\delta}
+\frac{1}{3}\Gamma^\sigma_{\tau\gamma}\Gamma^\tau_{\lambda\delta} \right].
\end{align}
\end{subequations}
The current $B^\mu$ is not really a vector under diffeomorphisms; it
transforms like a connection, but locally, this is irrelevant. In the
literature, sometimes the combination $W\equiv
R_\mn^2-R^2/3=\widehat{R}_\mn^2-R^2/12$ appears, so that
$G=C_{\kl\mn}^2-2W$. 

The global Gauss-Bonnet formula relates the integral of $R^*R^*$ to the Euler characteristic $\chi$ ,
\beq\label{eq:euler}
\int_M d^4x \sqrt{g} R^*R^*+\int_{\partial M} d^3x \sqrt{\gamma}B^\mu n_\mu=\,32\pi^2 \chi,
\eeq 
where it has been assumed that $M$ is an orientable, differentiable manifold in
four-dimensions, and ${\partial M}$ represents its possible boundaries.
($\gamma_\kl$ is the push-forward metric on the surface induced by $g_\mn,$ 
and $n^\mu$ is the outward pointing normal.)
The Euler number $\chi=2-2g$, where $g$ is the genus (number of
``handles".) The genus of the sphere $S^4$ is zero, so it has $\chi=2$.
This relation is very general and, with appropriate modifications of the
left-hand side, can be generalized to manifolds and non-smooth surfaces.
(It can even be defined topologically without reference to a metric.)

\section{Background Field Method}\label{sec:bfm}

\paragraph{ }
In this appendix, we review the background field method  very
briefly\footnote{Early reviews are presented in
Refs.~\cite{Abbott:1981ke, Barvinsky:1985an, DeWitt:1988dq}  and more
recent summaries in Refs.~\cite{Avramidi:2000pia,Buchbinder:1992rb}.}, 
since we need to refer to a few results in the text.  We shall employ
DeWitt's condensed notation~\cite{DeWitt:1965jb}, using a single index
to denote all indices, including spacetime $x^\mu$ or other continuous
parameters.  Repeated indices are (usually) summed or integrated over.

The effective action $\Gamma[\phi_i]=S[\phi_i]+\Delta\Gamma[\phi_i]$ includes all quantum corrections $\Delta\Gamma[\phi_i]$ to the classical action $S[\phi_i]$. $\Delta\Gamma[\phi_i]$ may be defined formally in terms of an integro-differential equation as follows, 
in a straightforward 
generalization of the original path integral treatment
of the effective potential~\cite{Jackiw:1974cv}. 
In the classical action, the fields of the theory are shifted $\phi_i\to\phi_i+h_i$, and the resulting change in the classical action beyond first order in $h_i$ is calculated:
\beq\label{eq:deltaclassical}
\Delta S[\phi_i,h_i]=S[\phi_i+h_i]-S[\phi_i]-h_j\frac{\delta S[\phi_i]}{\delta \phi_j}.
\eeq
Then one defines an auxiliary functional $\Delta \gcal[\phi_i;K_i]$ by
\beq\label{eq:kgen}
e^{-\Delta \gcal[\phi_i;K_i]}=\int_\bcal \dcal h_i  e^{-\Delta S[\phi_i,h_i] - h_k K^k}, 
\eeq
where $K^i$ is initially an arbitrary ``source function."  Then it can be shown that 
\begin{align}
&\frac{\delta \Delta\gcal[\phi_i;K_i]}{\delta K_j}=0,\ 
{\rm{when}}\ K^j[\phi_i] = -\frac{\delta \Delta\Gamma[\phi_i]}{\delta \phi_j},\ {\rm and,}
\label{eq:source}\\
&{\rm {for\ that\ value\ of} }\, K^j[\phi_i], \ \Delta\gcal[\phi_i;K_j[\phi_i]]=\Delta\Gamma[\phi_i].
\label{eq:effact}
\end{align}
The interpretation of the expression \eqn{eq:kgen} is that $ \Delta\gcal[\phi_i;K_i]
$ is the generating functional of 1PI Green's functions in $h_i$ for a
given background field $\phi_i.$  These can in principle be evaluated. 
Then the function $ \Delta\gcal[\phi_i;K_i] $ can be used to choose a
source function for a given $\phi_i$ so that the one-point function for
$h_i$ vanishes.  (Thus, the background $\phi_i$ is self-consistent.) 
For that source function $K_j[\phi_i]$, then $\Delta
\gcal[\phi_i; K_j[\phi_i]]=\Delta\Gamma[\phi_i]$, the quantum corrections
 to the classical action.

At first glance, this argument seems circular, 
but it is in fact well-suited to calculations in perturbation theory. 
By construction, when expanded in powers of $h_j$, 
the lowest order contribution to $\Delta S[\phi_i,h_i]$ is quadratic in $h_i$.  
This determines the propagator for $h_i$ as a function of the background field.  
To this order, it gives the well-known correction to the effective action equal to 
\beq\label{eq:effact1}
\Gamma^{(1)}[\phi_i]=
\frac{1}{2} \mathrm{Log\,Det} \left[\frac{\delta^2 S}{\delta\phi_i\delta\phi_j}\right]=
\frac{1}{2}\mathrm{Tr\,Log}\left[\frac{\delta^2 S}{\delta\phi_i\delta\phi_j}\right].
\eeq

There are numerous technical obstacles to implementing this machinery,
all of which have been overcome or circumvented.  The QFT defined by 
\eqns{eq:deltaclassical}{eq:source} generally requires renormalization,
so a cutoff must be introduced.   If the theory is renormalizable in the
traditional sense, then, in the simplest cases, the action $S$ will
contain a finite number of independent monomials in the fields along
with their associated coupling constants.  In more complicated cases,
such as gravity, it may contain an infinite number of terms whose
relation to each other is prescribed by a symmetry, \ie, the number of
coupling constants does not increase. If it is an effective
field theory, then $S$ will contain as many terms (and coupling constants) 
as are necessary in
order to achieve a given degree of accuracy. Regardless, the  fluctuations 
contributing to \eqn{eq:effact1}\ may include 
negative or zero eigenvalues.
Negative modes suggest either that the theory is ill-defined (such as
the flat-space $\phi^3$-model) or that the background chosen is not
self-consistent and must be modified.  Zero modes are ``flat directions"
in the space of fields $h_i$, which may be the result of a symmetry or
may be resolved by higher order terms in the expansion in $h_i$.  In any
case, it must be determined whether or not  $\Delta S[\phi_i,h_i]$ is
bounded from below as a function of $h_i$ or not.   In order to
interpret the ``classical" action in terms of renormalized fields and
couplings rather than ``bare" quantities, the fields and coupling
constants are rescaled in such a way as to render the quantum
corrections to Green's functions finite.  This is usually expressed by
saying that the action includes local counterterms chosen as functions
of the cutoff as needed.  This makes the determination of stability even
more difficult and provisional because it is insufficient to determine
simply that  $\Delta S[\phi_i,h_i]\ge0$ for bare fields and couplings
but must be true for the renormalized fields and couplings, which 
depend on the renormalisation scale.  Ultimately, it is the finite
effective action including quantum corrections that needs to be analyzed
to determine stability and, in some cases, such as the ones considered
in this paper, at certain scales the size of the quantum corrections can
be as large as the ``classical" corrections. As mentioned in the text,
one may even encounter instabilities at one scale that do not persist at
other scales.  Finally, since the effective action is nonlocal, the
criteria for the existence of a sensible background (``vacuum") and
stability is not so easily  established generically.

In gauge theories, one must introduce gauge-fixing terms in order to
obtain sensible Feynman rules, so the effective action is gauge
dependent except on-shell where $\delta\Gamma/\delta\phi_i=0.$  If the
gauge-fixing terms are cleverly chosen, one can maintain gauge
invariance of the effective action, but that does not mean that they are
independent of all gauge-fixing parameters.  
The
ambiguous choice of effective action can make the  determination of the
stability of a QFT off-shell in principle problematic. At
one-loop order, most definitions of a ``self-consistent"
background field do agree,  so AF models may not  suffer from such
ambiguities concerning their UV behavior.

\section{Global Scale Invariance}\label{sec:scaleinv}

\paragraph{ }
In this Appendix, we shall review how scale invariance in \eqn{eq:scaletransf} comes about.
The scaling symmetry is
\beq\label{eq:scaling}
x^\mu\to\widehat{x^\mu}=e^{-\alpha} x^\mu,\quad
\phi(x)\to \widehat{\phi}(\widehat{x})=e^\alpha\phi(x),
\quad {g_\mn}(x)\to \widehat{g_\mn}(\widehat{x})=
{g_\mn}(x),
\eeq
for arbitrary real $\alpha.$ Unlike with general coordinate
transformations, the invariant length is rescaled,
\beq
ds^2\!={g_\mn}(x)dx^\mu dx^\nu\! \to d\widehat{s}\,{}^2
\!=\widehat{g_\mn}(\widehat{x})\widehat{dx^\mu}\widehat{dx^\nu}\!=\exp(-2\alpha)ds^2.
\eeq
In contrast, diffeomorphism invariance corresponds to metric transformations leaving scalars
invariant and covariant lengths unchanged:
\beq
x\!\to{x}^\prime(x),\ \phi(x)\!\to \phi'(x')=\phi(x), \ 
ds^2\!=g_\mn(x)dx^\mu dx^\nu={g_\mn}{}^\prime(x') {dx^\mu}^\prime{dx^\nu}^\prime\!=ds'{}^2\!,
\eeq
where
\beq
{g_\mn}{}^\prime(x') = \frac{\pa{x}^{\lambda}}{\pa x^{'\mu}}
\frac{\pa{x}^{\sigma}}{\pa x^{'\nu}}g_{\lambda\sigma}(x).
\eeq
Thus if we make the scale transformation corresponding to \eqn{eq:scaling}, followed by the 
general coordinate transformation corresponding to 
\beq
\widehat{x^\mu} \to {x'}^{\mu} =e^{\alpha}\, \widehat{x^\mu},
\eeq
it is easy to see that we generate a transformation precisely of the form \eqn{eq:scaletransf}, 
with $x^{\prime} = x$.
Thus for a theory which is both scale invariant and general coordinate invariant, we can 
use this (more convenient) form. 

\section{Stability of One-Loop Effective Action} \label{sec:stability}

\paragraph{ }
In order to calculate the one-loop effective action using the background field method of Appendix~\ref{sec:bfm}, we must first form $\Delta S$, \eqn{eq:deltaclassical}.  We write the metric as 
$g_\mn\equiv g_\mn^B+h_\mn$ and the scalar field as 
$\phi=\phi_0+\delta\phi,$ where $g_\mn^B$ is the de~Sitter background metric in a convenient choice of coordinates associated with a constant curvature $R_0$, and $\phi_0$ is the background value of the scalar field.  In tree approximation, even though $R_0$ and $\phi_0$ are undetermined, their ratio is fixed to be $r=\phi_0^2/R_0=\xi(\mu)/\lambda(\mu).$  We shall assume this to be the case, \ie, we restrict our attention to the fluctuations on-shell in order to avoid discussing the complications associated with gauge-fixing.  
The fluctuations $h_\mn$ are  decomposed as~\cite{Fradkin:1981iu, Avramidi:2000pia}
\beq
h_\mn=h_\mn^\perp+\nabla_{\{\mu}\ecal_{\nu\}}+\frac{g_\mn^B}{4} \varphi,
\eeq
where $h_\mn^\perp$ is the spin-two projection of $h_\mn$ (traceless and transverse, 
$\nabla^\mu h_\mn^\perp=0,$) $\ecal_\mu$ is a four-vector, and $\varphi$ is a scalar.  
Indices are raised and lowered using the background metric $g_\mn^B$, and the implied  connection is with respect to the background metric. 
If we decompose $\ecal_\mu$ into its transverse (spin one) and longitudinal parts, 
$\ecal_\mu\equiv\ecal_\mu^\perp+\nabla_\mu\sigma/2,$ and define 
$h\equiv g^B{}^\mn h_\mn$, then we find that $\varphi=h-\Box\sigma$.  
 Under a gauge transformation, 
$\delta h_\mn=\nabla_{\{\mu}\Theta_{\nu\}}$, 
then $h_\mn^\perp$ and $\varphi$ are gauge invariant, and,
decomposing $\Theta_\mu\equiv \Theta_\mu^\perp+\nabla_\mu \Theta,$ 
$\delta\ecal_\mu^\perp=\Theta_\mu^\perp,$ and $\delta\sigma=2\Theta$, $\delta h=2\Box\Theta.$  We shall work ``on-shell" so that the gauge-dependent modes will not enter.  
For that matter, we could choose the ``unitary gauge" where $\sigma=0, \ecal_\mu^\perp=0.$

With this notation, then we find to second order in the fluctuations,
 \begin{subequations}
 \begin{align}\label{eq:jaction4}
\Delta S^{(2)}&\!=\!\!\int \! d^4x \sqrt{{g^B}}\! 
\left[\frac{1}{2} \Big(\delta\phi  \Delta_0\left(2\xi R_0\right)\delta\phi \Big)
-\delta\phi\frac{3\xi\phi_0}{4}
\Delta_0\left(-\frac{ R_0}{3}\right) \varphi 
+\delta^{(2)}\lcal_{ho} \right],\\
\delta^{(2)}\lcal_{ho}&= \left[ \frac{3}{16b}\varphi\Delta_0\left(-\frac{b\xi\phi_0^2}{4}\right)\Delta_0\left(-\frac{ R_0}{3}\right)\varphi +\right.\nonumber\\
&\left. \hskip10mm
+\frac{1}{4a}\overline{h}^{\,\perp}_\mn\Delta_2\left(\frac{a\xi\phi_0^2}{2}
+\frac{ R_0}{3}\left(1-2w\right)\right) \Delta_2\left(\frac{ R_0}{6}\right)\overline{h}^{\,\perp}{}^\mn
\right],
\end{align}
\end{subequations}
where $\Delta_j(X)\equiv-\Box_j+X$ acting on the constrained field of spin $j$. 
Expanding in eigenfunctions of the Laplacian on the sphere,
we have eigenvalues~\cite{Fradkin:1981iu}
\beq
\Box_j\equiv \rho_0^2\,\overline{\lambda}^{(j)}_n,~~\overline{\lambda}^{(j)}_n\equiv n(n+3)-j,\quad d_n^{(j)}\equiv \frac{2j+1}{6}(2n+3)[n(n+3)-j(j+1)+2],
\eeq
where $\rho_0^2\equiv R_0/12,$ $\overline{\lambda}^{(j)}_n$ is the eigenvalue on the unit sphere $S^4$, and $d_n^{(j)}$ is the degree of degeneracy of the eigenvalue $\overline{\lambda}^{(j)}_n.$
Then,  after integration over $S^4$, we find for the tensor modes\footnote{This is the same as the results in~\cite{Fradkin:1981iu,Avramidi:2000pia} with the appropriate assignment to their masses $m_0^2, m_2^2$.}
\begin{align}\label{eq:tensormodes}
\Delta S^{(2)}&=\frac{1}{4a}\sum_{n=2}^{\infty}d_n^{(2)}\left[ \frac{6\xi^2}{y}+4\left(1-2w\right)
+\overline{\lambda}^{(2)}_n\ \right]\left[ 2+ \overline{\lambda}^{(2)}_n\right]\left(\overline{h}_n^{\,\perp}\right)^2.
\end{align}
For convergence of integration over the large $n$-modes, we must have $a>0.$
There will be neither negative nor zero modes provided the $n=2$ mode, 
$\overline{h}_2^{\,\perp},$ has positive coefficient.  This requires 
$w<3/2+3\xi^2/(4y).$  We required   
$y>0$ for stability of the minimum at $r=r_0$. The couplings are to be evaluated at some
convenient scale $\mu$ where these inequalities are satisfied.

Returning to \eqn{eq:jaction4}, we find for the conformal scalar modes,
\begin{align}\label{eq:conformal}
\Delta S^{(2)}&=\frac{12\xi}{ay}\sum_{n=0}^\infty d_n^{(0)}
\left[\frac{1}{2} \left(24\xi+\overline{\lambda}^{(0)}_n\right)\left(\!\frac{\delta\phi_n}{\phi_0}\!\right)^{\!\!2} 
-\frac{3\xi}{4}\left(-4+\overline{\lambda}^{(0)}_n \right)
\left(\!\frac{\delta\phi_n}{\phi_0}\!\right) \varphi_n
+ \right.\cr
&\left. +\left(\frac{w y}{64\xi}\right)
\left(\!-\frac{3\xi^2}{w y}+\overline{\lambda}_n^{(0)} \!\right)
\left(-4+\overline{\lambda}_n^{(0)} \right) \varphi_n^2\right].
\end{align}
In order that the large-$n$ modes be positive, it is necessary that $w>0$ and $\xi>0.$

The $n=0$ mode has coefficient $d_0^{(0)}=1$ times
\begin{align}
 &\frac{36\xi^2}{ay}
\left[ 4\left(\!\frac{\delta\phi_0}{\phi_0}\!\right)^{\!\!2} 
+\left(\!\frac{\delta\phi_0}{\phi_0}\!\right) \varphi_0+\frac{1}{16}\varphi_0^2 \right].
\label{eq:scalarstability}
\end{align}
This mixing matrix has one positive eigenvalue $(585\xi^2/\lambda)$ and one zero eigenvalue, a flat direction.  The zero mode is easily understood.  Classically, scale invariance is broken by the background field, and, since this calculation of fluctuations represents simply an expansion of the classical action about a fixed background, there must be a Goldstone boson associated with spontaneous breaking of scale invariance.  Removing the zero mode, we get a contribution from the positive eigenvalue to the one-loop correction to the effective action.  The QFT {\it explicitly} breaks scale invariance owing to the running of the couplings, so we can hope that this zero mode is lifted in higher order, and indeed, at two loops, it obtains a contribution from $C^{(2)}\ne 0.$  This classical zero mode thus gets a mass as a result of the anomalous scale invariance.  Requiring its mass$^2$ to be positive gives a minimum of the action and removes the flat direction.

The next eigenvalue ($n=1$) has $\overline{\lambda}_1=4$ and $d_1^{(0)}=5$.  The fluctuations are then
\beq
\frac{12\xi}{ay}\left[2(6\xi+1)\left(\!\frac{\delta\phi_1}{\phi_0}\!\right)^{\!\!2}
\right].
\eeq
As in the pure gravity case, the coefficient of $\varphi_1^2$ vanishes, as well as the cross term $\varphi_1\delta\phi_1$.  Thus, we continue to find five zero modes associated with the vanishing of contributions from the $\varphi_1$ conformal mode.  The other eigenvalue is positive, $24\xi(6\xi+1)/(ay)>0$ for $\xi>0,$ which we require anyway in order that the gravitational constant $\xi\phi_0^2$ be positive.
  We do not understand the reason why these five zero modes persist, \ie, we do not understand this flat direction as the result of a symmetry, broken or unbroken, and we have the feeling that we may be missing something.  Unlike the dilaton mode, we do not know whether it remains flat in higher order or just what happens.

In sum, the constraints on the couplings in order that all modes be nonnegative are that all four couplings $a, w,\xi,y$ be positive at some scale and that 
$w<3/2+3\xi^2/(4y)$.

\section{One-Loop Beta-Functions}\label{sec:betas}

\paragraph{ }
We have taken results for $\beta_{a,b,\varepsilon}^{ho}$ from
\reference{Avramidi:1985ki},  which corrects earlier
results of Fradkin and Tseytlin
(\cite{Fradkin:1981hx},\cite{Fradkin:1981iu})  for $\beta_b$.
At one-loop order, the effect of matter on the gravitational 
beta-functions is simply to add another term, so that 
$\beta_a,$ $\beta_b$ (or $\beta_w,$) and 
$\beta_\varepsilon$ become sums
$\beta_i=\beta_i^{ho}+\beta_i^{mat}$.
These results also follow from gravitational trace anomaly calculations and
are well known:  see for example \reference{Birrell:1982ix}. For the gravitational
contributions to the matter $\beta$-functions $\beta_{\lambda,\xi},$ we
have used \reference{Buchbinder:1992rb} and references therein.
We summarise the $\beta$-functions below in our notation:


\begin{subequations}
\label{eq:betagravity}
\begin{align}
 &\frac{1}{\kappa}\beta_\varepsilon^{ho}\!=\!-\frac{196}{45},
 \hskip29mm
 \frac{1}{\kappa}\beta_\varepsilon^{mat}\!=\!-\frac{1}{360}\left[N_0\!+\!11N_{1/2}\!+\!62N_1^{0}\!+\!63N_1 \right];&\\
 &\frac{1}{\kappa}\beta_a^{ho}\!=\!-\frac{133}{10}a^2,
 \hskip26mm
 \frac{1}{\kappa}\beta_a^{mat}\!=\!-\frac{a^2}{60}\left[ N_0\!+\!6N_{1/2} \!+\!12N_1^{0}\!+\!13N_1\right];&\\
 &\frac{1}{\kappa}\beta_b^{ho}\!=\! -\frac{5}{3}\left[ 2a^2-3ab\!+\!\frac{b^2}{4} \right],
 \hskip3mm 
 \frac{1}{\kappa}\beta_b^{mat}\!=\!-\frac{b^2}{24}\left[ \left(1\!+\!6\xi\right)^2\!N_0\!+\!N_1 \right];  &\\
\begin{split}
&\frac{1}{\kappa}\beta_w^{ho}\! =\!\frac{10\,a}{3}\left[w^2-\frac{549}{100}\,w\!+\!\frac{1}{8} \right],\cr
&\hskip20mm \frac{1}{\kappa}\beta_w^{mat}\! =-\frac{a}{60}w\left[ N_0\!+\!6N_{1/2} \!+\!12N_1^{0}\!+\!13N_1\right]
\!+\!\frac{a}{24}\left[ \left(1\! +\!6\xi\right)^2\!N_0\!+\!N_1 \right]\!;
\end{split}
\end{align}
\end{subequations} 
where $N_0$ denotes the number of (real) scalars; $N_{1/2}$, DIRAC
fermions\footnote{Not, as stated in \reference{Avramidi:2000pia}, two-component ones.
};  $N_1^0$, massless vectors; $N_1$, massive vectors. (For chiral
or Majorana fermions,  the coefficients of $N_{1/2}$ would be half those
given above). The parameter $\xi$ represents the
non-minimal coupling of a real scalar; in general, one may have a sum of
such couplings. 
For a general theory 
we may write (at one loop)  $\beta_a=-\kappa \beta_2a^2,$  and 
$\beta_\varepsilon=-\kappa\beta_1$, with positive constants $\beta_2, \beta_1$. 
It can be shown~\cite{Einhorn:2014bka} that the leading contribution 
to $\varepsilon$ is determined to be  $\varepsilon=\varepsilon_0 -\beta_1/(\beta_2a)$   
where $\varepsilon_0$ is a scale-independent constant.

The one-loop beta-functions for the matter couplings obviously depend on the particular model. For the single, real scalar action \eqn{eq:jrealscalar} with couplings $\xi$ and $\lambda$, they are
\begin{subequations}
\label{eq:betamat}
\begin{align} 
&\frac{1}{\kappa}\beta_\xi^{ho}\!=\!
-a\xi\left[\frac{3\xi^2}{2}\!+\!4\xi-3\!+\!\frac{10\,w}{3}
-\frac{1}{w}\left(\frac{9\,\xi^2}{4} \!+\!5\xi-1\right) \right],
\hskip2mm
\frac{1}{\kappa}\beta_\xi^{mat}\!=\!\left(6\xi\!+\!1\right)\lambda;&\\ 
&\frac{1}{\kappa}\beta_\lambda^{ho}\!=\!\frac{a^2\xi^2}{2}
\left[5\!+\! \frac{(6\xi\!+\!1)^2}{4w^2} \right]\!
-\!a\lambda
\left[ 5\!+\!3\xi^2\!-\left(\frac{1\!+\!12\xi\!+\!33\xi^2}{2w} \right)\! \right]\!,\ 
\frac{1}{\kappa}\beta_\lambda^{mat}\!=\!18\lambda^2\!.\!&
\end{align}
\end{subequations}
This system of equations for $a,w,\xi,\lambda$ are rather complicated, 
but they can be somewhat simplified by introducing the variable 
$u\equiv (1/\beta_2)\log(a_0/a(\mu))$.  Then these equations may be
written as in \eqn{eq:betabarmat} in the text. 

\end{appendix}

\end{document}

%% file: mylatex.tex
\usepackage{bbm}                                                  

\def\nc{\newcommand}
\nc{\beq}{\begin{equation}}  \nc{\eeq}{\end{equation}}
\nc{\bea}{\begin{eqnarray}}  \nc{\eea}{\end{eqnarray}}
\nc{\bal}{\begin{align}}  \nc{\eal}{\end{align}}
\nc{\baa}{\begin{array}}     \nc{\eaa}{\end{array}}
\nc{\bit}{\begin{itemize}}   \nc{\eit}{\end{itemize}}
\nc{\ben}{\begin{enumerate}} \nc{\een}{\end{enumerate}}
\nc{\bce}{\begin{center}}    \nc{\ece}{\end{center}}
\nc{\bpm}{\begin{pmatrix}}   \nc{\epm}{\end{pmatrix}}
\nc{\bvt}{\begin{verbatim}}  \nc{\evt}{\end{verbatim}}
\def\half{\frac12}	

\def\to{\rightarrow}

\def\gesim{\gtrsim}
\def\lesim{\lesssim}
\def\gesim{\,{\raise-3pt\hbox{$\sim$}}\!\!\!\!\!{\raise2pt\hbox{$>$}}\,}
\def\lesim{\,{\raise-3pt\hbox{$\sim$}}\!\!\!\!\!{\raise2pt\hbox{$<$}}\,}
\def\boldoverdot{\,{\raise6pt\hbox{\bf.}\!\!\!\!\>}}

\def\ie{{\it i.e.}}
\def\eg{{$e.\,g.$}}

\def\bcal{{\cal B}}

\def\dcal{{\cal D}}
\def\ecal{{\cal E}}

\def\gcal{{\cal G}}

\def\lcal{{\cal L}}

\usepackage{bm}
\def\vev{vacuum expectation value}

\def\diag{\hbox{\diag}}

\def\ab{{\alpha\beta}}
\def\kl{{\kappa\lambda}}
\def\mn{{\mu\nu}}

\def\vevof#1{\left\langle #1 \right\rangle}

\def\doubleundertext#1{
{\undertext{\vphantom{y}#1}}\par\nobreak\vskip-\the\baselineskip\vskip4pt%
\undertext{\hbox to 2in{}}}
\def\inbox#1{\vbox{\hrule\hbox{\vrule\kern5pt
     \vbox{\kern5pt#1\kern5pt}\kern5pt\vrule}\hrule}}
\def\sqr#1#2{{\vcenter{\hrule height.#2pt
      \hbox{\vrule width.#2pt height#1pt \kern#1pt
         \vrule width.#2pt}
      \hrule height.#2pt} } }

\def\today{\ifcase\month\or
  January\or February\or March\or April\or May\or June\or
  July\or August\or September\or October\or November\or December\fi
  \space\number\day, \number\year}
\def\pmb#1{\setbox0=\hbox{#1}%
  \kern-.025em\copy0\kern-\wd0
  \kern.05em\copy0\kern-\wd0
  \kern-.025em\raise.0433em\box0 }

\def\pmbb#1{\setbox0=\hbox{#1}%
  \kern-.02em\copy0\kern-\wd0
  \kern.04em\copy0\kern-\wd0
  \kern-.02em\raise.03464em\box0 }
\def\sumprime_#1{\setbox0=\hbox{$\scriptstyle{#1}$}
  \setbox2=\hbox{$\displaystyle{\sum}$}
  \setbox4=\hbox{${}'\mathsurround=0pt$}
  \dimen0=.5\wd0 \advance\dimen0 by-.5\wd2
  \ifdim\dimen0>0pt
  \ifdim\dimen0>\wd4 \kern\wd4 \else\kern\dimen0\fi\fi
\mathop{{\sum}'}_{\kern-\wd4 #1}}